# Matrix genetics, part 4: cyclic changes of the genetic 8-dimensional Yin-Yang-algebras and the algebraic models of physiological cycles

Sergey V. Petoukhov
Department of Biomechanics, Mechanical Engineering Research Institute of the Russian Academy of Sciences
petoukhov@hotmail.com, petoukhov@imash.ru, http://symmetry.hu/isabm/petoukhov.html

**Corresponding address**: Kutuzovskiy prospect, d. 1/7, kv.58, Moscow, 121248, Russia.

**Abstract**. The article continues an analysis of the genetic 8-dimensional Yin-Yang-algebra. This algebra was revealed in a course of matrix researches of structures of the genetic code and it was described in the author's articles arXiv:0803.3330 and arXiv:0805.4692. The article presents data about many kinds of cyclic permutations of elements of the genetic code in the genetic (8x8)-matrix [C A; U G]$^{(3)}$ of 64 triplets, where C, A, U, G are letters of the genetic alphabet. These cyclic permutations lead to such reorganizations of the matrix form of presentation of the initial genetic Yin-Yang-algebra that arisen matrices serve as matrix forms of presentations of new Yin-Yang-algebras as well. They are connected algorithmically with Hadamard matrices. The discovered existence of a hierarchy of the cyclic changes of types of genetic Yin-Yang-algebras allows thinking about new algebraic-genetic models of cyclic processes in inherited biological systems including models of cyclic metamorphoses of animals. These cycles of changes of the genetic 8-dimensional algebras and of their 8-dimensional numeric systems have many analogies with famous facts and doctrines of modern and ancient physiology, medicine, etc. This viewpoint proposes that the famous idea by Pythagoras (about organization of natural systems in accordance with harmony of numerical systems) should be combined with the idea of cyclic changes of Yin-Yang-numeric systems in considered cases. This second idea reminds of the ancient idea of cyclic changes in nature. From such algebraic-genetic viewpoint, the notion of biological time can be considered as a factor of coordinating these hierarchical ensembles of cyclic changes of types of the genetic multi-dimensional algebras.

## 1 Introduction

The article continues an analysis of the genetic 8-dimensional Yin-Yang-algebra, which was described in the articles [Petoukhov, arXiv:0803.3330, arXiv:0805.4692] and in the book [Petoukhov, 2008]. This analysis allows revealing unknown properties of this genetic algebra and its possible applications for deeper understanding of genetic and physiological systems including inherited physiological cycles.

One of directions, where the results of analysis of Yin-Yang-algebras can be useful, is related with a creation of algebraic models of inherited physiological cycles and rhythms in organisms. The statement that biological organisms exist in accordance with cyclic processes of environment and with their own cyclic physiological processes is one of the most classical statements of biology and medicine from ancient times (see for example [Wright, 2002]). Many branches of medicine take into account the time of day specially, when diagnostic, pharmacological and therapeutic actions should be made for individuals. The set of this medical and biological knowledge is usually united under names of chrono-medicine and chrono-biology. Many diseases are connected with disturbances of natural biological rhythms in organisms. The problem of internal clocks of organisms, which participate in coordination of all interrelated processes of any organism, is one of the main physiological problems. But cyclic principles are essential for spatial organization of living bodies as well. Biological morphogenesis gives many examples of a cyclic symmetric repetition of separate spatial blocks in constructions of organism bodies (see, for example, [Petoukhov, 1989]).

Molecular biology deals with this problem of physiological rhythms and of cyclic re-combinations of molecular ensembles on the molecular level as well. Really, it is the well-known fact that in biological organisms proteins are disintegrated into amino acids and then they are re-built (are re-created) from amino acids again in a cyclic manner systematically. A half-life period (a duration of renovation of half of a set of molecules) for proteins of human organisms is approximately equal to 80 days in most cases; for proteins of the liver and blood plasma – 10 days; for the mucilaginous cover of bowels – 3-4 days; for insulin – 6-9 minutes. Such permanent rebuilding of proteins provides a permanent cyclic renovation of human organisms. These known facts are described in biological encyclopedias (for example, see [Aksenova, 1998, v. 2, p. 19]). Such cyclic processes at the molecular-genetic level are one of the parts of a hierarchical system of a huge number of interelated cycles in organisms. The phenomenon of repeated recombinations of molecular ensembles, which are carried out inside the separate cycles, is one of the main problems of biological self-organization. This phenomenon takes additional attention to structural properties of recombinations and permutations of molecular elements of genetic code systems. We are studying these structural properties on a matrix language.

Whether some structural connections of the genetic code systems with inherited physiological rhythms and with such cyclic processes exist? Matrix genetics proposes new mathematical data of structural analysis for a positive answer on the first question and for a creation of algebraic models of such hierarchical system of cyclic changes. These data were received on the base of an analysis of the mentioned genetic 8-dimensional Yin-Yang-algebra. This algebra was revealed initially in the result of analysis of the genetic matrix $[C\ \ A;\ U\ \ G]^{(3)}$, where the symbol in parentheses means the third Kronecker power and the symbols C, A, U, G mark nitrogenous bases of the genetic code (cytosine, adenine, uracil, guanine).

This genetic algebra defines the system of 8-dimensional numbers YY with 8 real coordinates $x_0$, $x_1$, …, $x_7$ (the matrix form of presentation of these numbers YY is shown on Figures 1, 2):

$$YY = x_0*\mathbf{f_0}+x_1*\mathbf{m_1}+x_2*\mathbf{f_2}+x_3*\mathbf{m_3}+x_4*\mathbf{f_4}+x_5*\mathbf{m_5}+x_6*\mathbf{f_6}+x_7*\mathbf{m_7} \qquad (1)$$

Multiplication of any two members of such octet numbers YY generates a new octet number of the same system. This situation is similar to the situation of real numbers (or of complex numbers, or of hypercomplex numbers) when multiplication of any two members of a numeric system generates a new member of the same numeric system. From the abstract mathematical viewpoint such numeric system can be used for modeling not only static systems but of variable systems and processes as well. For such case of variable processes, one should consider coordinates of YY-numbers in the expression (1) as variable functions of time: $x_0(t)$, $x_1(t)$, $x_2(t)$, $x_3(t)$, $x_4(t)$, $x_5(t)$, $x_6(t)$, $x_7(t)$. For example, these functions can be trigonometric functions like sine(w*t) and cosine(w*t) or they can be Walsh-Hadamard functions, etc. Multiplication of any two such 8-dimensional superpositions of functions gives a new 8-dimensional superposition of functions, which corresponds to the expression (1) again. In special cases these variable functions $x_0(t)$, $x_1(t)$, ..., $x_7(t)$ can describe some permutations of elements in systems with variable compositions, etc.

| | | | | | | | |
|---|---|---|---|---|---|---|---|
| **CCC**<br>**Pro, $x_0$** | **CCA**<br>**Pro, $x_1$** | **CAC**<br>**His, $-x_2$** | **CAA**<br>**Gln, $-x_3$** | **ACC**<br>**Thr, $x_4$** | **ACA**<br>**Thr, $x_5$** | **AAC**<br>**Asn, $-x_6$** | **AAA**<br>**Lys, $-x_7$** |
| **CCU**<br>**Pro, $x_0$** | **CCG**<br>**Pro, $x_1$** | **CAU**<br>**His, $-x_2$** | **CAG**<br>**Gln, $-x_3$** | **ACU**<br>**Thr, $x_4$** | **ACG**<br>**Thr, $x_5$** | **AAU**<br>**Asn, $-x_6$** | **AAG**<br>**Lys, $-x_7$** |
| **CUC**<br>**Leu, $x_2$** | **CUA**<br>**Leu, $x_3$** | **CGC**<br>**Arg, $x_0$** | **CGA**<br>**Arg, $x_1$** | **AUC**<br>**Ile, $-x_6$** | **AUA**<br>**Met, $-x_7$** | **AGC**<br>**Ser, $-x_4$** | **AGA**<br>**Stop, $-x_5$** |
| **CUU**<br>**Leu, $x_2$** | **CUG**<br>**Leu, $x_3$** | **CGU**<br>**Arg, $x_0$** | **CGG**<br>**Arg, $x_1$** | **AUU**<br>**Ile, $-x_6$** | **AUG**<br>**Met, $-x_7$** | **AGU**<br>**Ser, $-x_4$** | **AGG**<br>**Stop, $-x_5$** |
| **UCC**<br>**Ser, $x_4$** | **UCA**<br>**Ser, $x_5$** | **UAC**<br>**Tyr, $-x_6$** | **UAA**<br>**Stop, $-x_7$** | **GCC**<br>**Ala, $x_0$** | **GCA**<br>**Ala, $x_1$** | **GAC**<br>**Asp, $-x_2$** | **GAA**<br>**Glu, $-x_3$** |
| **UCU**<br>**Ser, $x_4$** | **UCG**<br>**Ser, $x_5$** | **UAU**<br>**Tyr, $-x_6$** | **UAG**<br>**Stop, $-x_7$** | **GCU**<br>**Ala, $x_0$** | **GCG**<br>**Ala , $x_1$** | **GAU**<br>**Asp, $-x_2$** | **GAG**<br>**Glu, $-x_3$** |
| **UUC**<br>**Phe, $-x_6$** | **UUA**<br>**Leu, $-x_7$** | **UGC**<br>**Cys, $-x_4$** | **UGA**<br>**Trp, $-x_5$** | **GUC**<br>**Val, $x_2$** | **GUA**<br>**Val, $x_3$** | **GGC**<br>**Gly, $x_0$** | **GGA**<br>**Gly, $x_1$** |
| **UUU**<br>**Phe, $-x_6$** | **UUG**<br>**Leu, $-x_7$** | **UGU**<br>**Cys, $-x_4$** | **UGG**<br>**Trp, $-x_5$** | **GUU**<br>**Val, $x_2$** | **GUG**<br>**Val, $x_3$** | **GGU**<br>**Gly, $x_0$** | **GGG**<br>**Gly, $x_1$** |

*Figure 1. The genetic matrix [C A; U G]$^{(3)}$ of 64 triplets, 20 amino acids and 4 stop-codons for the vertebrate mitochondrial genetic code. The black-and-white mosaic of the matrix reflects the degeneracy of the code (see [Petoukhov, arXiv:0802.3366]). The Yin-coordinates $x_0$, $x_2$, $x_4$, $x_6$ and the Yang-coordinates $x_1$, $x_3$, $x_5$, $x_7$ of the 8-dimensional Yin-Yang-algebra $YY_+^{[CAUG]}$ are shown additionally (from [Petoukhov, arXiv:0803.3330]). The black cells contain the coordinates with the sign „+" and the white cells contain the coordinates with the sign „-".*

Figure 1 shows the genetic matrix [C A; U G]$^{(3)}$ of 64 triplets for the case of the vertebrate mitochondria genetic code, which is considered as the basic dialect of the genetic code. This genetic matrix (or "genomatrix" briefly) possesses the black-and-white mosaic, which reflects the specifics of the degeneracy of the code and which has special symmetric properties. This genetic matrix is transformed (by means of the "alphabetic" algorithm of the Yin-Yang-digitization of 64 triplets) into the matrix form of presentation of 8-dimensional numbers YY from the expression (1) (see details in [Petoukhov, arXiv:0803.3330]).

If the matrix on Figure 1 has the eight real coordinates $x_0$, $x_1$, …, $x_7$ in its cells only, we have the matrix $YY_+^{[CAUG]}$ (Figure 2, on the right side) as the matrix form of presentation of a new numeric system of 8-dimensional numbers YY from the expression (1) ([Petoukhov, arXiv:0803.3330]). The upper index in the symbol $YY_+^{[CAUG]}$ shows the kind of symbolic genomatrices, which is transformed into this kind of numeric Yin-Yang-matrices by means of the mentioned algorithm of the Yin-Yang-digitization of 64 triplets. A meaning of the lower index "+" will be explained in the second half of the article. Below we will meet with many kinds of Yin-Yang-algebras, which are produced by means of cyclic permutations of genetic elements in the genetic matrix [C A; U G]$^{(3)}$ (Figure 1).

The multiplication table of this 8-dimensional numeric system $YY_+^{[CAUG]}$ (or its algebra over a field) is shown on the left side of Figure 2. Cells of the main diagonal of this multiplication table contain squares of the basic elements only. In cases of usual hypercomplex numbers these diagonal cells contain real units "±1" typically. But in our case these diagonal cells contain no real units at all but all diagonal cells are occupied by elements "$\pm\mathbf{f_0}$" and "$\pm\mathbf{m_1}$". Thereby the set of the 8 basic matrices $\mathbf{f_0}$, $\mathbf{m_1}$, $\mathbf{f_2}$, $\mathbf{m_3}$, $\mathbf{f_4}$, $\mathbf{m_5}$, $\mathbf{f_6}$, $\mathbf{m_7}$ is divided into two equal subsets by criterion of their squares. The first subset consists of elements with the even indexes: $\mathbf{f_0}$, $\mathbf{f_2}$, $\mathbf{f_4}$, $\mathbf{f_6}$. The squares of members of this $\mathbf{f_0}$-subset are equal to $\pm\mathbf{f_0}$ always. The second subset consists of elements with the odd indexes: $\mathbf{m_1}$, $\mathbf{m_3}$, $\mathbf{m_5}$, $\mathbf{m_7}$. The squares of members of this $\mathbf{m_1}$-subset are equal to $\pm\mathbf{m_1}$ always.

| | №0 | №1 | №2 | №3 | №4 | №5 | №6 | №7 |
|---|---|---|---|---|---|---|---|---|
| | $f_0$ | $m_1$ | $f_2$ | $m_3$ | $f_4$ | $m_5$ | $f_6$ | $m_7$ |
| $f_0$ | $f_0$ | $m_1$ | $f_2$ | $m_3$ | $f_4$ | $m_5$ | $f_6$ | $m_7$ |
| $m_1$ | $f_0$ | $m_1$ | $f_2$ | m3 | $f_4$ | $m_5$ | $f_6$ | $m_7$ |
| $f_2$ | $f_2$ | $m_3$ | $-f_0$ | $-m_1$ | $-f_6$ | $-m_7$ | $f_4$ | $m_5$ |
| $m_3$ | $f_2$ | $m_3$ | $-f_0$ | $-m_1$ | $-f_6$ | $-m_7$ | $f_4$ | $m_5$ |
| $f_4$ | $f_4$ | $m_5$ | $f_6$ | $m_7$ | $f_0$ | $m_1$ | $f_2$ | $m_3$ |
| $m_5$ | $f_4$ | $m_5$ | $f_6$ | $m_7$ | $f_0$ | $m_1$ | $f_2$ | $m_3$ |
| $f_6$ | $f_6$ | $m_7$ | $-f_4$ | $-m_5$ | $-f_2$ | $-m_3$ | $f_0$ | $m_1$ |
| $m_7$ | $f_6$ | $m_7$ | $-f_4$ | $-m_5$ | $-f_2$ | $-m_3$ | $f_0$ | $m_1$ |

| №0 | №1 | №2 | №3 | №4 | №5 | №6 | №7 |
|---|---|---|---|---|---|---|---|
| $x_0$ | $x_1$ | $-x_2$ | $-x_3$ | $x_4$ | $x_5$ | $-x_6$ | $-x_7$ |
| $x_0$ | $x_1$ | $-x_2$ | $-x_3$ | $x_4$ | $x_5$ | $-x_6$ | $-x_7$ |
| $x_2$ | $x_3$ | $x_0$ | $x_1$ | $-x_6$ | $-x_7$ | $-x_4$ | $-x_5$ |
| $x_2$ | $x_3$ | $x_0$ | $x_1$ | $-x_6$ | $-x_7$ | $-x_4$ | $-x_5$ |
| $x_4$ | $x_5$ | $-x_6$ | $-x_7$ | $x_0$ | $x_1$ | $-x_2$ | $-x_3$ |
| $x_4$ | $x_5$ | $-x_6$ | $-x_7$ | $x_0$ | $x_1$ | $-x_2$ | $-x_3$ |
| $-x_6$ | $-x_7$ | $-x_4$ | $-x_5$ | $x_2$ | $x_3$ | $x_0$ | $x_1$ |
| $-x_6$ | $-x_7$ | $-x_4$ | $-x_5$ | $x_2$ | $x_3$ | $x_0$ | $x_1$ |

*Figure 2. On the left: the table of multiplication of the basic elements $f_0$, $m_1$, $f_2$, $m_3$, $f_4$, $m_5$, $f_6$, $m_7$ of the octet algebra $YY_+^{[CAUG]}$. On the right: the matrix form of presentation of the algebra $YY_+^{[CAUG]}$ (the Yin-Yang-numeric presentation of the genomatrix $[C\ A;\ U\ G]^{(3)}$ from Figure 1).*

The basic element $\mathbf{f_0}$ possesses all properties of the real unit in relation to the members of the $\mathbf{f_0}$-subset: $\mathbf{f_0}^2 = \mathbf{f_0}$, $\mathbf{f_0*f_2=f_2*f_0=f_2}$, $\mathbf{f_0*f_4=f_4*f_0=f_4}$, $\mathbf{f_0*f_6=f_6*f_0=f_6}$. But the element $\mathbf{f_0}$ does not possess the commutative property of real unity in relation to the members of the $\mathbf{m_1}$-subset: $\mathbf{f_0*m_p} \neq \mathbf{m_p*f_0}$, where p=1, 3, 5, 7. By this reason $\mathbf{f_0}$ is named "quasi-real unity of the $\mathbf{f_0}$-subset".

The basic element $\mathbf{m_1}$ possesses all properties of the real unit in relation to the members of the $\mathbf{m_1}$-subset: $\mathbf{m_1}^2\mathbf{=m_1}$, $\mathbf{m_1*m_3=m_3*m_1=m_3}$, $\mathbf{m_1*m_5=m_5*m_1=m_5}$, $\mathbf{m_1*m_7=m_7*m_1=m_7}$. But the element $\mathbf{m_1}$ does not possess the commutative property of real unity in relation to the members of the $\mathbf{f_0}$-subset: $\mathbf{m_1*f_k} \neq \mathbf{f_k*m_1}$, where k=0, 2, 4, 6**.** By this reason $\mathbf{m_1}$ is named "quasi-real unity of the $\mathbf{m_1}$-subset".

The principle "even-odd" exists in this algebra $YY_+^{[CAUG]}$. Really all members of the $\mathbf{f_0}$-subset and their coordinates $x_0$, $x_2$, $x_4$, $x_6$ have even indexes and they are disposed in columns with the even numbers 0, 2, 4, 6 in the matrix $YY_+^{[CAUG]}$ and in its multiplication table also. All members of the $\mathbf{m_1}$-subset and their coordinates $x_1$, $x_3$, $x_5$, $x_7$ have the odd indexes and they are disposed in columns with the even numbers 1, 3, 5, 7 in the matrix $YY_+^{[CAUG]}$ and in its multiplication table (Figure 2) as well. In accordance with Pythagorean and Ancient-Chinese traditions, all even numbers are named "female" numbers or Yin-numbers, and all odd numbers are named "male" numbers or Yang-numbers. From the viewpoint of this tradition, the elements $\mathbf{f_0}$, $\mathbf{f_2}$, $\mathbf{f_4}$, $\mathbf{f_6}$, $x_0$, $x_2$, $x_4$, $x_6$ with the even indexes play the role of "female" elements or Yin-elements, and the elements $\mathbf{m_1}$, $\mathbf{m_3}$, $\mathbf{m_5}$, $\mathbf{m_7}$, $x_1$, $x_3$, $x_5$, $x_7$ with the odd indexes play the role of "male" or Yang-elements. Correspondingly this 8-dimensional algebra $YY_+^{[CAUG]}$ was named as the octet Yin-Yang-algebra (or the even-odd-algebra, or the bisex-algebra). This Yin-Yang-algebra was utilized as a useful model of the genetic code and of evolution of its dialects [Petoukhov, arXiv:0805.4692].

## 2 Revealing new Yin-Yang-algebras in the result of cyclic permutations of genetic elements

Matrix genetics studies structures of the genetic code by means of matrix methods of the theory of discrete signals processing. This theory pays a great attention to permutations of discrete elements in information processing [Ahmed, Rao, 1975]. In view of this, the following question arises: what genetic matrices are produced by various cyclic permutations of genetic elements in the initial genetic matrices [C A; U G]$^{(3)}$? Whether these new matrices possess interesting mathematical properties?

One of unexpected results of studying this question is the discovery that many kinds of such cyclic permutations generate new genetic matrices, which are related algorithmically with matrices of new kinds of 8-dimensional Yin-Yang-algebras as well. The transformation of these new genetic matrices into matrices of new Yin-Yang-algebras is carried out by means of the same alphabetic algorithm of the Yin-Yang-digitization of 64 triplets (see a description of this algorithm in [Petoukhov, arXiv:0803.3330]). This algorithm connects solidly each genetic triplet with one of the eight YY-coordinates $x_1$, $x_2$, $x_3$, $x_4$, $x_5$, $x_6$, $x_7$ (with its certain sign "+" or "-" in accordance with Figure 1) and this connection is irrespective of a disposition of triplets in considered genetic (8x8)-matrices of 64 triplets.

We begin with the case of a circular permutation of the genetic letters C→A→G→U→C. This circular permutation means that in all triplets of the genomatrix [C A; U G]$^{(3)}$ the letter A is replaced by the letter C, the letter G is replaced by the letter A, etc. In the result a new genomatrix [U C; G A]$^{(3)}$ arises, which is shown on Figure 3.

This symbolic genomatrix defines the appropriate numeric Yin-Yang-matrix $YY_+^{[UCGA]}$ by means of the algorithm of the Yin-Yang-digitization of 64 triplets. Numeric components of this matrix $YY_+^{[UCGA]}$ are disposed in the matrix cells on Figure 3. One can see that this new genomatrix [U C; G A]$^{(3)}$ can be received by another way at all. Really, the same matrix arises when the initial matrix [C A; U G]$^{(3)}$ (Figure 1) is turned on $90^0$ clockwise.

| | | | | | | | |
|---|---|---|---|---|---|---|---|
| UUU<br>Phe, $-x_6$ | UUC<br>Phe, $-x_6$ | UCU<br>Ser, $x_4$ | UCC<br>Ser, $x_4$ | CUU<br>Leu, $x_2$ | CUC<br>Leu, $x_2$ | CCU<br>Pro, $x_0$ | CCC<br>Pro, $x_0$ |
| UUG<br>Leu, $-x_7$ | UUA<br>Leu, $-x_7$ | UCG<br>Ser, $x_5$ | UCA<br>Ser, $x_5$ | CUG<br>Leu, $x_3$ | CUA<br>Leu, $x_3$ | CCG<br>Pro, $x_1$ | CCA<br>Pro, $x_1$ |
| UGU<br>Cys, $-x_4$ | UGC<br>Cys, $-x_4$ | UAU<br>Tyr, $-x_6$ | UAC<br>Tyr, $-x_6$ | CGU<br>Arg, $x_0$ | CGC<br>Arg, $x_0$ | CAU<br>His, $-x_2$ | CAC<br>His, $-x_2$ |
| UGG<br>Trp, $-x_5$ | UGA<br>Trp, $-x_5$ | UAG<br>Stop, $-x_7$ | UAA<br>Stop, $-x_7$ | CGG<br>Arg, $x_1$ | CGA<br>Arg, $x_1$ | CAG<br>Gln, $-x_3$ | CAA<br>Gln, $-x_3$ |
| GUU<br>Val, $x_2$ | GUC<br>Val, $x_2$ | GCU<br>Ala, $x_0$ | GCC<br>Ala, $x_0$ | AUU<br>Ile, $-x_6$ | AUC<br>Ile, $-x_6$ | ACU<br>Thr, $x_4$ | ACC<br>Thr, $x_4$ |
| GUG<br>Val, $x_3$ | GUA<br>Val, $x_3$ | GCG<br>Ala , $x_1$ | GCA<br>Ala, $x_1$ | AUG<br>Met, $-x_7$ | AUA<br>Met, $-x_7$ | ACG<br>Thr, $x_5$ | ACA<br>Thr, $x_5$ |
| GGU<br>Gly, $x_0$ | GGC<br>Gly, $x_0$ | GAU<br>Asp, $-x_2$ | GAC<br>Asp, $-x_2$ | AGU<br>Ser, $-x_4$ | AGC<br>Ser, $-x_4$ | AAU<br>Asn, $-x_6$ | AAC<br>Asn, $-x_6$ |
| GGG<br>Gly, $x_1$ | GGA<br>Gly, $x_1$ | GAG<br>Glu, $-x_3$ | GAA<br>Glu, $-x_3$ | AGG<br>Stop, $-x_5$ | AGA<br>Stop, $-x_5$ | AAG<br>Lys, $-x_7$ | AAA<br>Lys, $-x_7$ |

*Figure 3. The genetic matrix [U C; G A]$^{(3)}$, which arises from the genetic matrix [C A; U G]$^{(3)}$ (Figure 1) in the result of the circular permutation of the genetic letters C→A→G→U→C. All designations are the same as on Figure 1.*

If the same circular permutation C→A→G→U→C is repeated for the new genomatrix [U C; G A]$^{(3)}$, the genomatrix [G U; A C]$^{(3)}$ arises (Figure 4). The same matrix [G U; A C]$^{(3)}$ can be received by means of a turn of the matrix [U C; G A]$^{(3)}$ on $90^0$ clockwise as well.

| | | | | | | | |
|---|---|---|---|---|---|---|---|
| GGG<br>Gly, $x_1$ | GGU<br>Gly, $x_0$ | GUG<br>Val, $x_3$ | GUU<br>Val, $x_2$ | UGG<br>Trp, $-x_5$ | UGU<br>Cys, $-x_4$ | UUG<br>Leu, $-x_7$ | UUU<br>Phe, $-x_6$ |
| GGA<br>Gly, $x_1$ | GGC<br>Gly, $x_0$ | GUA<br>Val, $x_3$ | GUC<br>Val, $x_2$ | UGA<br>Trp, $-x_5$ | UGC<br>Cys, $-x_4$ | UUA<br>Leu, $-x_7$ | UUC<br>Phe, $-x_6$ |
| GAG<br>Glu, $-x_3$ | GAU<br>Asp, $-x_2$ | GCG<br>Ala , $x_1$ | GCU<br>Ala, $x_0$ | UAG<br>Stop, $-x_7$ | UAU<br>Tyr, $-x_6$ | UCG<br>Ser, $x_5$ | UCU<br>Ser, $x_4$ |
| GAA<br>Glu, $-x_3$ | GAC<br>Asp, $-x_2$ | GCA<br>Ala, $x_1$ | GCC<br>Ala, $x_0$ | UAA<br>Stop, $-x_7$ | UAC<br>Tyr, $-x_6$ | UCA<br>Ser, $x_5$ | UCC<br>Ser, $x_4$ |
| AGG<br>Stop, $-x_5$ | AGU<br>Ser, $-x_4$ | AUG<br>Met, $-x_7$ | AUU<br>Ile, $-x_6$ | CGG<br>Arg, $x_1$ | CGU<br>Arg, $x_0$ | CUG<br>Leu, $x_3$ | CUU<br>Leu, $x_2$ |
| AGA<br>Stop, $-x_5$ | AGC<br>Ser, $-x_4$ | AUA<br>Met, $-x_7$ | AUC<br>Ile, $-x_6$ | CGA<br>Arg, $x_1$ | CGC<br>Arg, $x_0$ | CUA<br>Leu, $x_3$ | CUC<br>Leu, $x_2$ |
| AAG<br>Lys, $-x_7$ | AAU<br>Asn, $-x_6$ | ACG<br>Thr, $x_5$ | ACU<br>Thr, $x_4$ | CAG<br>Gln, $-x_3$ | CAU<br>His, $-x_2$ | CCG<br>Pro, $x_1$ | CCU<br>Pro, $x_0$ |
| AAA<br>Lys, $-x_7$ | AAC<br>Asn, $-x_6$ | ACA<br>Thr, $x_5$ | ACC<br>Thr, $x_4$ | CAA<br>Gln, $-x_3$ | CAC<br>His, $-x_2$ | CCA<br>Pro, $x_1$ | CCC<br>Pro, $x_0$ |

*Figure 4. The genetic matrix* [G U; A C]$^{(3)}$, *which arises from the initial genetic matrix [C A; U G]$^{(3)}$ (Figure 1) in the result of the twice applications of the circular permutation of the genetic letters* C→A→G→U→C. *All designations are the same as on Figure 1.*

If the same circular permutation C→A→G→U→C is repeated for the new genomatrix [G U; A C]$^{(3)}$, the genomatrix [A G; C U]$^{(3)}$ arises (Figure 5). The same matrix [A G; C U]$^{(3)}$ can be received by means of a turn of the matrix [G U; A C]$^{(3)}$ on $90^0$ clockwise as well.

| | | | | | | | |
|---|---|---|---|---|---|---|---|
| AAA<br>Lys, $-x_7$ | AAG<br>Lys, $-x_7$ | AGA<br>Stop, $-x_5$ | AGG<br>Stop, $-x_5$ | GAA<br>Glu, $-x_3$ | GAG<br>Glu, $-x_3$ | GGA<br>Gly, $x_1$ | GGG<br>Gly, $x_1$ |
| AAC<br>Asn, $-x_6$ | AAU<br>Asn, $-x_6$ | AGC<br>Ser, $-x_4$ | AGU<br>Ser, $-x_4$ | GAC<br>Asp, $-x_2$ | GAU<br>Asp, $-x_2$ | GGC<br>Gly, $x_0$ | GGU<br>Gly, $x_0$ |
| ACA<br>Thr, $x_5$ | ACG<br>Thr, $x_5$ | AUA<br>Met, $-x_7$ | AUG<br>Met, $-x_7$ | GCA<br>Ala, $x_1$ | GCG<br>Ala , $x_1$ | GUA<br>Val, $x_3$ | GUG<br>Val, $x_3$ |
| ACC<br>Thr, $x_4$ | ACU<br>Thr, $x_4$ | AUC<br>Ile, $-x_6$ | AUU<br>Ile, $-x_6$ | GCC<br>Ala, $x_0$ | GCU<br>Ala, $x_0$ | GUC<br>Val, $x_2$ | GUU<br>Val, $x_2$ |
| CAA<br>Gln, $-x_3$ | CAG<br>Gln, $-x_3$ | CGA<br>Arg, $x_1$ | CGG<br>Arg, $x_1$ | UAA<br>Stop, $-x_7$ | UAG<br>Stop, $-x_7$ | UGA<br>Trp, $-x_5$ | UGG<br>Trp, $-x_5$ |
| CAC<br>His, $-x_2$ | CAU<br>His, $-x_2$ | CGC<br>Arg, $x_0$ | CGU<br>Arg, $x_0$ | UAC<br>Tyr, $-x_6$ | UAU<br>Tyr, $-x_6$ | UGC<br>Cys, $-x_4$ | UGU<br>Cys, $-x_4$ |
| CCA<br>Pro, $x_1$ | CCG<br>Pro, $x_1$ | CUA<br>Leu, $x_3$ | CUG<br>Leu, $x_3$ | UCA<br>Ser, $x_5$ | UCG<br>Ser, $x_5$ | UUA<br>Leu, $-x_7$ | UUG<br>Leu, $-x_7$ |
| CCC<br>Pro, $x_0$ | CCU<br>Pro, $x_0$ | CUC<br>Leu, $x_2$ | CUU<br>Leu, $x_2$ | UCC<br>Ser, $x_4$ | UUC<br>Phe, $-x_6$ | UUC<br>Phe, $-x_6$ | UUU<br>Phe, $-x_6$ |

*Figure 5. The genetic matrix [A G; C U]$^{(3)}$, which arises from the initial genetic matrix [C A; U G]$^{(3)}$ (Figure 1) in the result of the triple applications of the circular permutation of the genetic letters* C→A→G→U→C. *All designations are the same as on Figure 1.*

Dispositions of the 8 coordinates $x_1$, $x_2$, …, $x_7$ are different in these four symbolic genomatrices [C A; U G]$^{(3)}$, [U C; G A]$^{(3)}$, [G U; A C]$^{(3)}$ and [A G; C U]$^{(3)}$. They define the four different numeric matrices $YY_+^{[CAUG]}$, $YY_+^{[UCGA]}$, $YY_+^{[GUAC]}$, $YY_+^{[AGCU]}$ by the same algorithm correspondingly The beautiful fact is that the new three numeric matrices $YY_+^{[CAUG]}$, $YY_+^{[UCGA]}$, $YY_+^{[GUAC]}$, $YY_+^{[AGCU]}$ present appropriate 8-dimensional Yin-Yang-algebras as well by analogy with the matrix $YY_+^{[CAUG]}$ on Figure 2. For example, let us consider the numeric matrix $YY_+^{[UCGA]}$, which is reproduced on Figure 6.

| | | | | | | | |
|---|---|---|---|---|---|---|---|
| $-x_6$ | $-x_6$ | $x_4$ | $x_4$ | $x_2$ | $x_2$ | $x_0$ | $x_0$ |
| $-x_7$ | $-x_7$ | $x_5$ | $x_5$ | $x_3$ | $x_3$ | $x_1$ | $x_1$ |
| $-x_4$ | $-x_4$ | $-x_6$ | $-x_6$ | $x_0$ | $x_0$ | $-x_2$ | $-x_2$ |
| $-x_5$ | $-x_5$ | $-x_7$ | $-x_7$ | $x_1$ | $x_1$ | $-x_3$ | $-x_3$ |
| $x_2$ | $x_2$ | $x_0$ | $x_0$ | $-x_6$ | $-x_6$ | $x_4$ | $x_4$ |
| $x_3$ | $x_3$ | $x_1$ | $x_1$ | $-x_7$ | $-x_7$ | $x_5$ | $x_5$ |
| $x_0$ | $x_0$ | $-x_2$ | $-x_2$ | $-x_4$ | $-x_4$ | $-x_6$ | $-x_6$ |
| $x_1$ | $x_1$ | $-x_3$ | $-x_3$ | $-x_5$ | $-x_5$ | $-x_7$ | $-x_7$ |

*Figure 6. The numeric matrix $YY_+^{[UCGA]}$, which is numeric presentation of the genomatrix $[U\ \ C;\ G\ \ A]^{(3)}$ from Figure 3.*

This matrix $YY_+^{[UCGA]}$ can be written in the linear form in accordance with the expression (1): $YY = x_0*\mathbf{f_0}+x_1*\mathbf{m_1}+x_2*\mathbf{f_2}+x_3*\mathbf{m_3}+x_4*\mathbf{f_4}+x_5*\mathbf{m_5}+x_6*\mathbf{f_6}+x_7*\mathbf{m_7}$. The basic matrices $\mathbf{f_0}$, $\mathbf{m_1}$, $\mathbf{f_2}$, $\mathbf{m_3}$, $\mathbf{f_4}$, $\mathbf{m_5}$, $\mathbf{f_6}$, $\mathbf{m_7}$ for this case are shown on Figure 7.

This set of the basic matrices is the closed set relative to multiplication: the result of multiplication of any two basic matrices is a matrix from the same set in accordance with the multiplication table on Figure 8.

This multiplication table defines an 8-dimensional algebra again. The diagonal cells of this table contain no real units at all but they are occupied by elements "$\pm\mathbf{f_6}$" and "$\pm\mathbf{m_7}$". Thereby the set of the 8 basic matrices $\mathbf{f_0}$, $\mathbf{m_1}$, $\mathbf{f_2}$, $\mathbf{m_3}$, $\mathbf{f_4}$, $\mathbf{m_5}$, $\mathbf{f_6}$, $\mathbf{m_7}$ is divided into two equal subsets by criterion of their squares. The first subset consists of elements with the even indexes: $\mathbf{f_0}$, $\mathbf{f_2}$, $\mathbf{f_4}$, $\mathbf{f_6}$. The squares of members of this $\mathbf{f_6}$-subset are equal to $\pm\mathbf{f_6}$ always. The second subset consists of elements with the odd indexes: $\mathbf{m_1}$, $\mathbf{m_3}$, $\mathbf{m_5}$, $\mathbf{m_7}$. The squares of members of this $\mathbf{m_7}$-subset are equal to $\pm\mathbf{m_7}$ always.

The basic element $\mathbf{f_6}$ possesses all properties of real negative unit "-1" in relation to the members of the $\mathbf{f_6}$-subset: $\mathbf{f_6}^2 = -\mathbf{f_6}$, $\mathbf{f_6}*\mathbf{f_0} = \mathbf{f_0}*\mathbf{f_6} = -f_0$, $f_6*\mathbf{f_2} = \mathbf{f_2}*\mathbf{f_6} = -\mathbf{f_2}$, $\mathbf{f_6}*\mathbf{f_4} = \mathbf{f_4}*\mathbf{f_6} = -\mathbf{f_4}$. But the element $\mathbf{f_6}$ does not possess the commutative property of real negative unit in relation to the members of the $\mathbf{m_7}$-subset: $\mathbf{f_6}*\mathbf{m_p} \neq \mathbf{m_p}*\mathbf{f_6}$, where p=1, 3, 5, 7. By this reason $\mathbf{f_6}$ is named "quasi-real negative unit of the $\mathbf{f_6}$-subset".

The basic element $\mathbf{m_7}$ possesses all properties of real negative unit "-1" in relation to the members of the $\mathbf{m_7}$-subset: $\mathbf{m_7}^2 = -\mathbf{m_7}$, $\mathbf{m_7}*\mathbf{m_1} = \mathbf{m_1}*\mathbf{m_7} = -\mathbf{m_1}$, $\mathbf{m_7}*\mathbf{m_3} = \mathbf{m_3}*\mathbf{m_7} = -\mathbf{m_3}$, $\mathbf{m_7}*\mathbf{m_5} = \mathbf{m_5}*\mathbf{m_7} = -\mathbf{m_5}$. But the element $\mathbf{m_7}$ does not possess the commutative property of real negative unit in relation to the members of the $\mathbf{f_6}$-subset: $\mathbf{m_7}*\mathbf{f_k} \neq \mathbf{f_k}*\mathbf{m_7}$, where k=0, 2, 4, 6**.** By this reason $\mathbf{m_7}$ is named "quasi-real negative unit of the $\mathbf{m_7}$-subset".

$$f_0 = \begin{bmatrix} 0&0&0&0&0&0&1&1 \\ 0&0&0&0&0&0&0&0 \\ 0&0&0&0&1&1&0&0 \\ 0&0&0&0&0&0&0&0 \\ 0&0&1&1&0&0&0&0 \\ 0&0&0&0&0&0&0&0 \\ 1&1&0&0&0&0&0&0 \\ 0&0&0&0&0&0&0&0 \end{bmatrix} ; \quad m_1 = \begin{bmatrix} 0&0&0&0&0&0&0&0 \\ 0&0&0&0&0&0&1&1 \\ 0&0&0&0&0&0&0&0 \\ 0&0&0&0&1&1&0&0 \\ 0&0&0&0&0&0&0&0 \\ 0&0&1&1&0&0&0&0 \\ 0&0&0&0&0&0&0&0 \\ 1&1&0&0&0&0&0&0 \end{bmatrix}$$

$$f_2 = \begin{bmatrix} 0&0&0&0&1&1&0&0 \\ 0&0&0&0&0&0&0&0 \\ 0&0&0&0&0&0&-1&-1 \\ 0&0&0&0&0&0&0&0 \\ 1&1&0&0&0&0&0&0 \\ 0&0&0&0&0&0&0&0 \\ 0&0&-1&-1&0&0&0&0 \\ 0&0&0&0&0&0&0&0 \end{bmatrix} ; \quad m_3 = \begin{bmatrix} 0&0&0&0&0&0&0&0 \\ 0&0&0&0&1&1&0&0 \\ 0&0&0&0&0&0&0&0 \\ 0&0&0&0&0&0&-1&-1 \\ 0&0&0&0&0&0&0&0 \\ 1&1&0&0&0&0&0&0 \\ 0&0&0&0&0&0&0&0 \\ 0&0&-1&-1&0&0&0&0 \end{bmatrix}$$

$$f_4 = \begin{bmatrix} 0&0&1&1&0&0&0&0 \\ 0&0&0&0&0&0&0&0 \\ -1&-1&0&0&0&0&0&0 \\ 0&0&0&0&0&0&0&0 \\ 0&0&0&0&0&0&1&1 \\ 0&0&0&0&0&0&0&0 \\ 0&0&0&0&-1&-1&0&0 \\ 0&0&0&0&0&0&0&0 \end{bmatrix} ; \quad m_5 = \begin{bmatrix} 0&0&0&0&0&0&0&0 \\ 0&0&1&1&0&0&0&0 \\ 0&0&0&0&0&0&0&0 \\ -1&-1&0&0&0&0&0&0 \\ 0&0&0&0&0&0&0&0 \\ 0&0&0&0&0&0&1&1 \\ 0&0&0&0&0&0&0&0 \\ 0&0&0&0&-1&-1&0&0] \end{bmatrix}$$

$$f_6 = \begin{bmatrix} -1&-1&0&0&0&0&0&0 \\ 0&0&0&0&0&0&0&0 \\ 0&0&-1&-1&0&0&0&0 \\ 0&0&0&0&0&0&0&0 \\ 0&0&0&0&-1&-1&0&0 \\ 0&0&0&0&0&0&0&0 \\ 0&0&0&0&0&0&-1&-1 \\ 0&0&0&0&0&0&0&0 \end{bmatrix} ; \quad m_7 = \begin{bmatrix} 0&0&0&0&0&0&0&0 \\ -1&-1&0&0&0&0&0&0 \\ 0&0&0&0&0&0&0&0 \\ 0&0&-1&-1&0&0&0&0 \\ 0&0&0&0&0&0&0&0 \\ 0&0&0&0&-1&-1&0&0 \\ 0&0&0&0&0&0&0&0 \\ 0&0&0&0&0&0&-1&-1 \end{bmatrix}$$

*Figure 7. The basic matrices for the numeric matrix* $YY_+^{[UCGA]}$ *from Figure 6*

| | $f_0$ | $m_1$ | $f_2$ | $m_3$ | $f_4$ | $m_5$ | $f_6$ | $m_7$ |
|---|---|---|---|---|---|---|---|---|
| $f_0$ | $-f_6$ | $-f_6$ | $-f_4$ | $-f_4$ | $-f_2$ | $-f_2$ | $-f_0$ | $-f_0$ |
| $m_1$ | $-m_7$ | $-m_7$ | $-m_5$ | $-m_5$ | $-m_3$ | $-m_3$ | $-m_1$ | $-m_1$ |
| $f_2$ | $f_4$ | $f_4$ | $-f_6$ | $-f_6$ | $f_0$ | $f_0$ | $-f_2$ | $-f_2$ |
| $m_3$ | $m_5$ | $m_5$ | $-m_7$ | $-m_7$ | $m_1$ | $m_1$ | $-m_3$ | $-m_3$ |
| $f_4$ | $f_2$ | $f_2$ | $-f_0$ | $-f_0$ | $f_6$ | $f_6$ | $-f_4$ | $-f_4$ |
| $m_5$ | $m_3$ | $m_3$ | $-m_1$ | $-m_1$ | $m_7$ | $m_7$ | $-m_5$ | $-m_5$ |
| $f_6$ | $-f_0$ | $-f_0$ | $-f_2$ | $-f_2$ | $-f_4$ | $-f_4$ | $-f_6$ | $-f_6$ |
| $m_7$ | $-m_1$ | $-m_1$ | $-m_3$ | $-m_3$ | $-m_5$ | $-m_5$ | $-m_7$ | $-m_7$ |

*Figure 8. The multiplication table of the basic matrices of the numeric matrix* $YY_+^{[UCGA]}$ *from Figure 6.*

By definition, an Yin-Yang-algebra is a $2^n$-dimensional algebra, a complete set of basic elements of which has not real unit at all but this set consists of two sub-sets of basic elements with $2^{n-1}$ elements in each and with the following feature: one of basic elements of each sub-set possesses all properties of real positive unit "+1" or of real negative unit "-1" relative to all basic elements of its sub-set but not relative to basic elements of another sub-set. One can see that the multiplication table on Figure 8 defines the 8-dimensional Yin-Yang-algebra $YY_+^{[UCGA]}$. This genetic algebra for the case of the genomatrix $[U\ \ C;\ G\ \ A]^{(3)}$ is quite different from the genetic Yin-Yang-algebra $YY_+^{[CAUG]}$ with the multiplication table on Figure 2.

This difference can be illustrated additionally by a numeric example. Let us consider an arbitrary 8-dimensional number:

$$YY = 3*\mathbf{f_0}+4*\mathbf{m_1}+7*\mathbf{f_2}+2*\mathbf{m_3}+2*\mathbf{f_4}+1*\mathbf{m_5}+9*\mathbf{f_6}+8*\mathbf{m_7}. \qquad (2)$$

If this number is considered from the viewpoint of the genetic Yin-Yang-algebra $YY_+^{[CAUG]}$ on Figure 2, its square is equal to $YY^2 = 117*\mathbf{f_0} + 149*\mathbf{m_1} + 69*\mathbf{f_2} + 57*\mathbf{m_3} - 15*\mathbf{f_4} + 57*\mathbf{m_5} + 117*\mathbf{f_6} + 121*\mathbf{m_7}$. But if this number (2) with the same magnitudes of the coordinates $x_0, x_1, \ldots, x_7$ is considered from the viewpoint of the genetic Yin-Yang-algebra $YY_+^{[UCGA]}$ on Figure 8, its square is equal to the quite another number: $YY^2 = -111*\mathbf{f_0} - 127x_1*\mathbf{m_1} - 195*\mathbf{f_2} - 111*\mathbf{m_3} - 39*\mathbf{f_4} - 63*\mathbf{m_5} - 231*\mathbf{f_6} - 179*\mathbf{m_7}$.

One can note a certain similarity (or a symmetric relation) between the multiplication tables of the both cases on Figures 2 and 8. Really, the internal contents of the first table can be transformed into the internal contents of the second table by means of a turn on $90^0$ anticlockwise with simultaneous inversion of all signs "+" and "-". Symmetric relations exist also among multiplication tables of many genetic Yin-Yan-matrices, which are mentioned below.

For the cases of other two matrices $YY_+^{[GUAC]}$ and $YY_+^{[AGCU]}$, which correspond to the genomatrices $[G\ \ U;\ A\ \ C]^{(3)}$ (Figure 4) and $[A\ \ G;\ C\ \ U]^{(3)}$ (Figure 5), their appropriate sets of basic matrices and multiplication tables are constructed by analogy. Figures 9 and 10 demonstrate final results in a form of their multiplication tables. One can see that these two multiplication tables define two 8-dimensional Yin-Yang-algebras as well.

| | $f_0$ | $m_1$ | $f_2$ | $m_3$ | $f_4$ | $m_5$ | $f_6$ | $m_7$ |
|---|---|---|---|---|---|---|---|---|
| $f_0$ | $f_0$ | $m_1$ | $f_2$ | $m_3$ | $f_4$ | $m_5$ | $f_6$ | $m_7$ |
| $m_1$ | $f_0$ | $m_1$ | $f_2$ | $m_3$ | $f_4$ | $m_5$ | $f_6$ | $m_7$ |
| $f_2$ | $f_2$ | $m_3$ | $-f_0$ | $-m_1$ | $-f_6$ | $-m_7$ | $f_4$ | $m_5$ |
| $m_3$ | $f_2$ | $m_3$ | $-f_0$ | $-m_1$ | $-f_6$ | $-m_7$ | $f_4$ | $m_5$ |
| $f_4$ | $f_4$ | $m_5$ | $f_6$ | $m_7$ | $f_0$ | $m_1$ | $f_2$ | $m_3$ |
| $m_5$ | $f_4$ | $m_5$ | $f_6$ | $m_7$ | $f_0$ | $m_1$ | $f_2$ | $m_3$ |
| $f_6$ | $f_6$ | $m_7$ | $-f_4$ | $-m_5$ | $-f_2$ | $-m_3$ | $f_0$ | $m_1$ |
| $m_7$ | $f_6$ | $m_7$ | $-f_4$ | $-m_5$ | $-f_2$ | $-m_3$ | $f_0$ | $m_1$ |

*Figure 9. The multiplication table of the Yin-Yang-algebra $YY_+^{[GUAC]}$*

| | $f_0$ | $m_1$ | $f_2$ | $m_3$ | $f_4$ | $m_5$ | $f_6$ | $m_7$ |
|---|---|---|---|---|---|---|---|---|
| $f_0$ | $-f_6$ | $-f_6$ | $-f_4$ | $-f_4$ | $-f_2$ | $-f_2$ | $-f_0$ | $-f_0$ |
| $m_1$ | $-m_7$ | $-m_7$ | $-m_5$ | $-m_5$ | $-m_3$ | $-m_3$ | $-m_1$ | $-m_1$ |
| $f_2$ | $f_4$ | $f_4$ | $-f_6$ | $-f_6$ | $f_0$ | $f_0$ | $-f_2$ | $-f_2$ |
| $m_3$ | $m_5$ | $m_5$ | $-m_7$ | $-m_7$ | $m_1$ | $m_1$ | $-m_3$ | $-m_3$ |
| $f_4$ | $f_2$ | $f_2$ | $-f_0$ | $-f_0$ | $f_6$ | $f_6$ | $-f_4$ | $-f_4$ |
| $m_5$ | $m_3$ | $m_3$ | $-m_1$ | $-m_1$ | $m_7$ | $m_7$ | $-m_5$ | $-m_5$ |
| $f_6$ | $-f_0$ | $-f_0$ | $-f_2$ | $-f_2$ | $-f_4$ | $-f_4$ | $-f_6$ | $-f_6$ |
| $m_7$ | $-m_1$ | $-m_1$ | $-m_3$ | $-m_3$ | $-m_5$ | $-m_5$ | $-m_7$ | $-m_7$ |

*Figure 10. The multiplication table of the Yin-Yang-algebra $YY_+^{[AGCU]}$*

The four set of basic matrices $f_0$, $m_1$, $f_2$, $m_3$, $f_4$, $m_5$, $f_6$, $m_7$ for the four matrix $YY_+^{[CAUG]}$, $YY_+^{[UCGA]}$, $YY_+^{[GUAC]}$ and $YY_+^{[AGCU]}$ are quite different, but the multiplication tables for the matrices $YY_+^{[CAUG]}$ and $YY_+^{[GUAC]}$ (Figures 2 and 9) are identical to each other. These matrices are transformed each into another by means of turn on $180^0$ or by means of simultaneous permutations of complementary nitrogenous bases C↔G and A↔U in all triplets in the appropriate genomatrices $[C\ \ A;\ U\ \ G]^{(3)}$ and $[G\ \ U;\ A\ \ C]^{(3)}$ (that is each codon is replaced by its anti-codon in this case). It means, that for the case of these "complementary" genomatrices $[C\ \ A;\ U\ \ G]^{(3)}$ and $[G\ \ U;\ A\ \ C]^{(3)}$ the same Yin-Yang-algebra possesses two different matrix forms of its presentation. The similar situation is held true for the second pair of the "complementary" genomatrices $[U\ \ C;\ G\ \ A]^{(3)}$ and $[A\ \ G;\ C\ \ U]^{(3)}$: their Yin-Yang-algebras $YY_+^{[UCGA]}$ and $YY_+^{[AGCU]}$ are identical to each another as well because of identity of their multiplication tables (Figures 8 and 10).

One should add that each of these four genomatrices $[C\ \ A;\ U\ \ G]^{(3)}$, $[G\ \ U;\ A\ \ C]^{(3)}$, $[U\ \ C;\ G\ \ A]^{(3)}$ and $[A\ \ G;\ C\ \ U]^{(3)}$ corresponds to its own Hadamard matrix by means of a general alphabetic algorithm. The speech is that each of these (8x8)-matrices possesses a black-and-white mosaic, which is transformed easily into a mosaic of an appropriate Hadamard (8x8)-matrix by means of the U-algorithm. This U-algorithm is described in [Petoukhov, arXiv:0802.3366] and it is based on objective molecular properties of the genetic alphabet. These four "genetic" Hadamard matrices, which are marked by $H_+^{[CAUG]}$, $H_+^{[GUAC]}$, $H_+^{[UCGA]}$, $H_+^{[AGCU]}$, are shown on Figure 11.

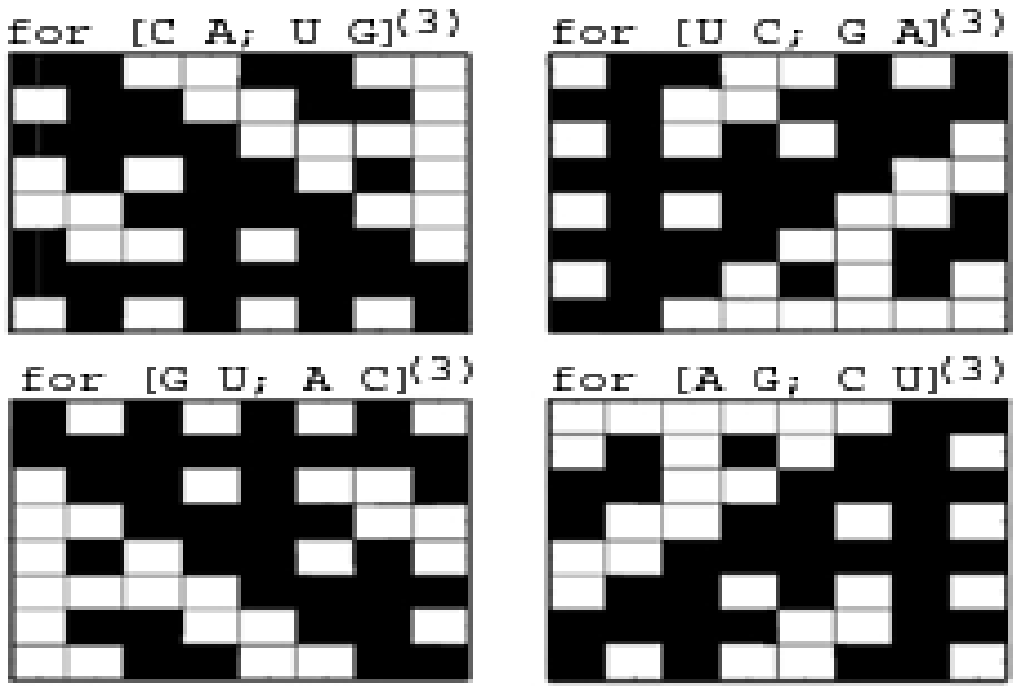

*Figure 11. The four Hadamard matrices $H_+^{[CAUG]}$, $H_+^{[GUAC]}$, $H_+^{[UCGA]}$, $H_+^{[AGCU]}$, which are connected algorithmically with the genomatrices $[C\ \ A;\ U\ \ G]^{(3)}$, $[U\ \ C;\ G\ \ A]^{(3)}$, $[G\ \ U;\ A\ \ C]^{(3)}$, $[A\ \ G;\ C\ \ U]^{(3)}$ and their Yin-Yang-algebras. Each black (white) cell contains the element "+1" ("-1").*

Each of the four genomatrices $[C\ \ A;\ U\ \ G]^{(3)}$, $[U\ \ C;\ G\ \ A]^{(3)}$, $[G\ \ U;\ A\ \ C]^{(3)}$ and $[A\ \ G;\ C\ \ U]^{(3)}$ can be transformed by a circular manner into the next genomatrix of this sequence by means of a turn on $90^0$ clockwise. Correspondingly each of the four Yin-Yang-matrices $YY_+^{[CAUG]}$, $YY_+^{[UCGA]}$, $YY_+^{[GUAC]}$ and $YY_+^{[AGCU]}$ and each of the four Hadamard matrices $H_+^{[CAUG]}$, $H_+^{[GUAC]}$, $H_+^{[UCGA]}$, $H_+^{[AGCU]}$ can be transformed by a circular manner into the next matrix of their sequences by means of a turn on $90^0$ clockwise. It is important result that the cyclic permutations of genetic elements C→A→G→U→C lead to the appropriate cyclic changes of the Yin-Yang-algebras $YY_+^{[CAUG]} \rightarrow YY_+^{[UCGA]} \rightarrow YY_+^{[GUAC]} \rightarrow YY_+^{[AGCU]} \rightarrow YY_+^{[CAUG]}$ in the matrix forms of presentation of the genetic code. These cyclic permutations lead simultaneously to the appropriate cyclic changes of the genomatrices $[C\ \ A;\ U\ \ G]^{(3)} \rightarrow [U\ \ C;\ G\ \ A]^{(3)} \rightarrow [G\ \ U;\ A\ \ C]^{(3)} \rightarrow [A\ \ G;\ C\ \ U]^{(3)} \rightarrow [C\ \ A;\ U\ \ G]^{(3)}$ and of the Hadamard matrices $H_+^{[CAUG]} \rightarrow H_+^{[GUAC]} \rightarrow H_+^{[UCGA]} \rightarrow H_+^{[AGCU]} \rightarrow H_+^{[CAUG]}$. Figure 12 illustrates this general situation for the sequence of the Yin-Yang-matrices only. Symbols of a clock and of the four parts of the world are disposed in the center of Figure 12 to cause heuristic associations and to reflect the thought that the spatial turns of these matrices (together with changes of their Yin-Yang-algebras and of their Hadamard matrices) can be carried out rhythmically in appropriate algebraic models of rhythmic physiological processes.

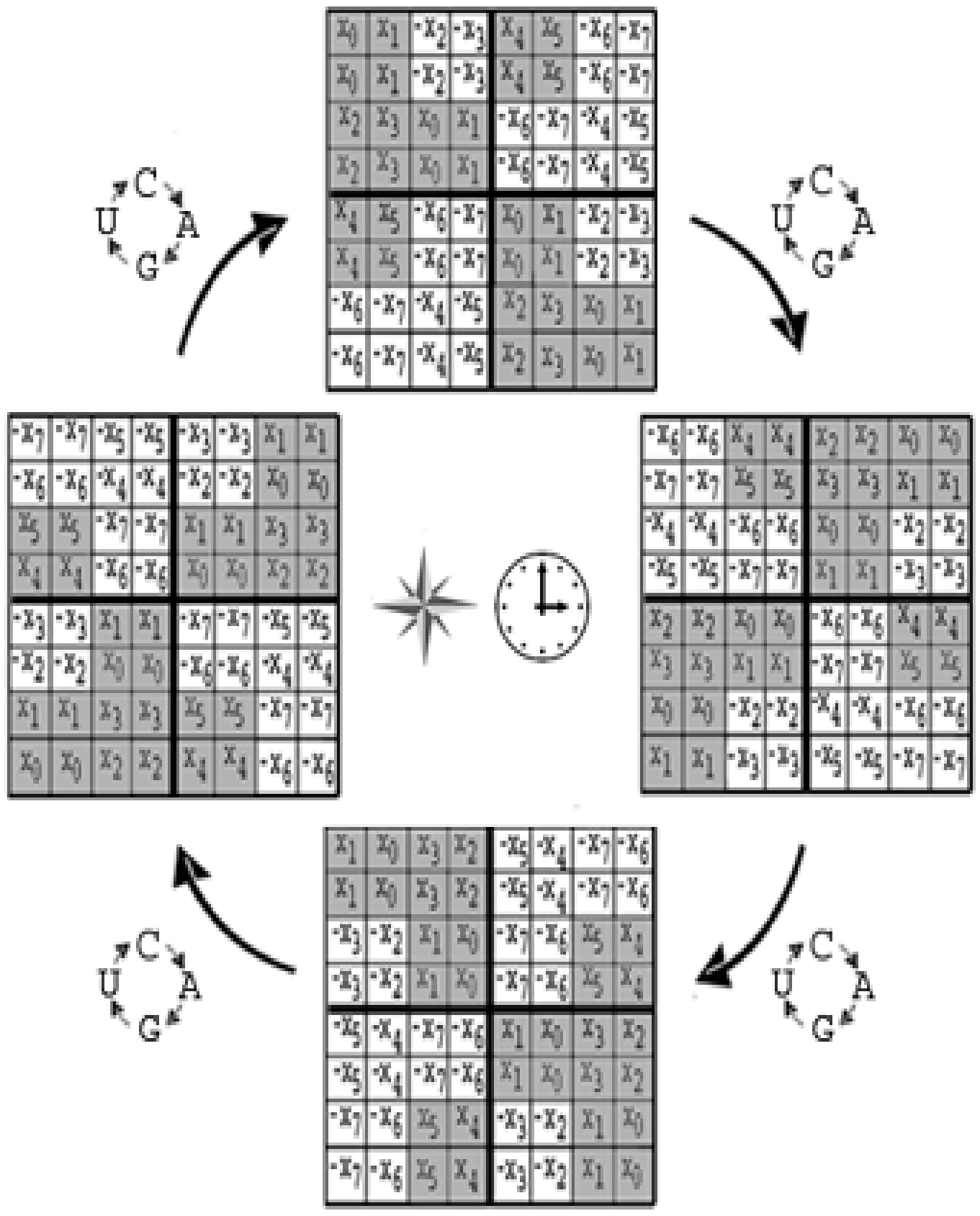

*Figure 12. The circular sequence of the four Yin-Yang-matrices $YY_+^{[CAUG]}$ → $YY_+^{[UCGA]}$ → $YY_+^{[GUAC]}$ → $YY_+^{[AGCU]}$ → $YY_+^{[CAUG]}$, which is based on the circular permutations of the genetic molecular elements C→A→G→U→C.*

But these circular sequences of the genetic Yin-Yang-algebras and of the conjunct matrices are a small part of a hierarchy of circular changes of different kinds of the genetic Yin-Yang-algebras together with conjunct matrices. This hierarchy, which is based on different kinds of circular permutations, should be studied step by step. Let us continue this study.

We begin with the genomatrix $[C\ \ A;\ U\ \ G]^{(3)}$ and the cyclic shifts 1-2-3→2-3-1→3-1-2→1-2-3 of three positions in all triplets there (for example, in the case of the triplet CAG such shifts produce the sequence CAG→AGC→GCA→CAG). Analogical cyclic shifts of the three positions 3-2-1→2-1-3→1-3-2→3-2-1 in the triplets at the reverse order of their reading are possible for analysis. Such cyclic changes produce the sequences of appropriate genomatrices $[C\ \ A;\ U\ \ G]_{123}^{(3)}$→ $[C\ \ A;\ U\ \ G]_{231}^{(3)}$→$[C\ \ A;\ U\ \ G]_{312}^{(3)}$ →$[C\ \ A;\ U\ \ G]_{123}^{(3)}$ and $[C\ \ A;\ U\ \ G]_{321}^{(3)}$→$[C\ \ A;\ U\ \ G]_{213}^{(3)}$→ $[C\ \ A;\ U\ \ G]_{132}^{(3)}$→$[C\ \ A;\ U\ \ G]_{321}^{(3)}$. Each of these six genomatrices is connected with its own Yin-Yang-algebras by the same algorithm (see details in the article [Petoukhov, arXiv:0803.3330] and the book [Petoukhov, 2008]). Figure 13 shows one of possible variants of schematic illustration of these two cyclic sequences of the mosaic Yin-Yang-matrices $YY_{+,123}^{[CAUG]}$, $YY_{+,231}^{[CAUG]}$, $YY_{+,312}^{[CAUG]}$, $YY_{+,321}^{[CAUG]}$, $YY_{+,213}^{[CAUG]}$, $YY_{+,132}^{[CAUG]}$. The multiplication tables of these six relevant algebras] are shown in the article [Petoukhov, arXiv:0803.3330, Figures 5, 8-12].

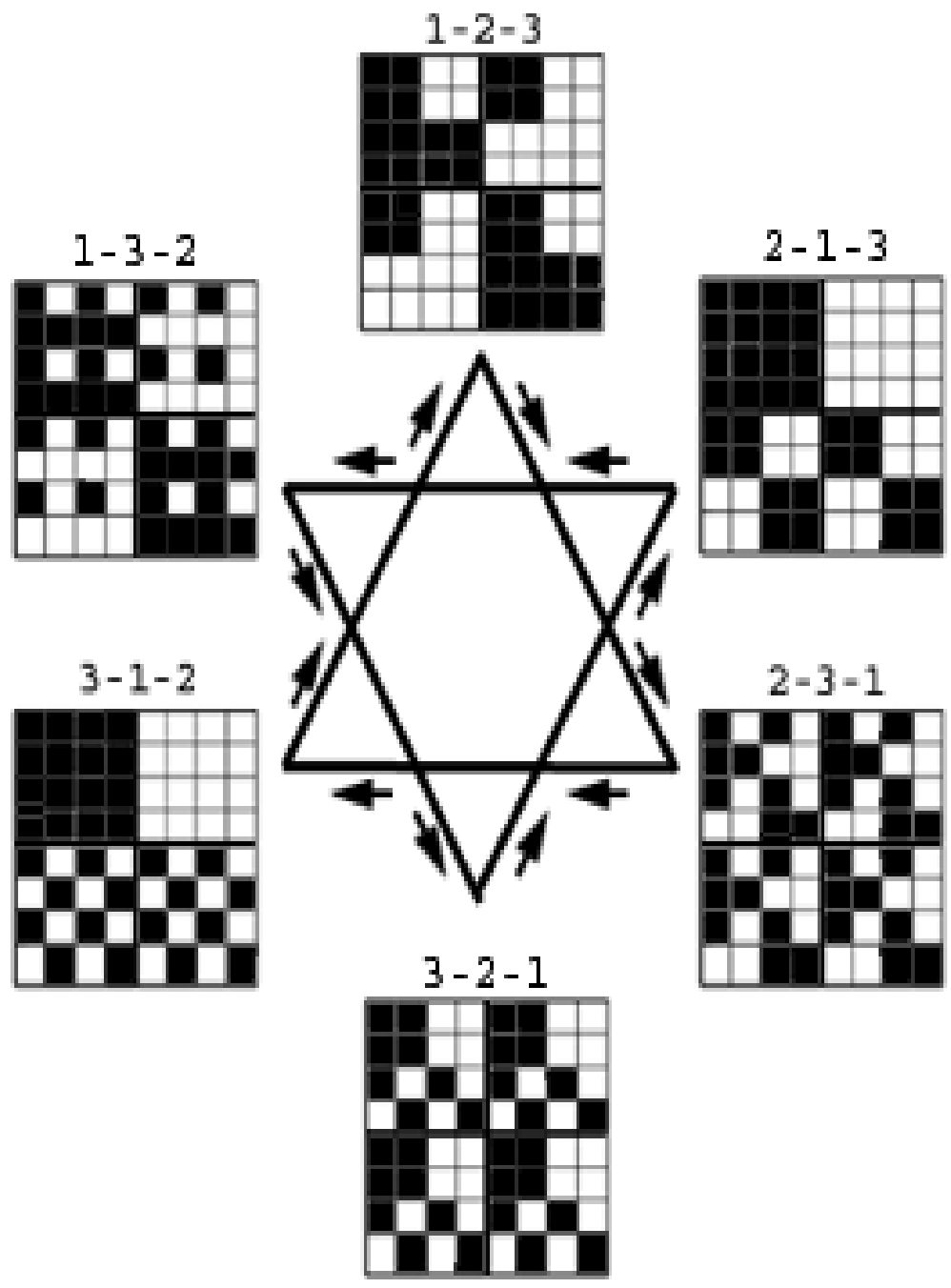

*Figure 13. A schematic illustration of the two cyclic sequences of the mosaic Yin-Yang-matrices, which arise in the result of permutations of positions in triplets:* $YY_{+,123}^{[CAUG]} \rightarrow YY_{+,231}^{[CAUG]} \rightarrow YY_{+,312}^{[CAUG]} \rightarrow YY_{+,123}^{[CAUG]}$ *and* $YY_{+,321}^{[CAUG]} \rightarrow YY_{+,213}^{[CAUG]} \rightarrow YY_{+,132}^{[CAUG]} \rightarrow YY_{+,321}^{[CAUG]}$. *Number over each matrix shows a relevant kind of permutations of positions in all triplets.*

A transposition of these Yin-Yang-matrices produces new relevant kinds of Yin-Yang-matrices [Petoukhov, 2008]. Each of these 12 Yin-Yang-matrices corresponds to its own kind of genetic Hadamard (8x8)-matrices.

By analogy the similar consideration of permutations and transpositions in the cases of other three initial genomatrices $[U\ \ C;\ G\ \ A]^{(3)}$, $[G\ \ U;\ A\ \ C]^{(3)}$ and $[A\ \ G;\ C\ \ U]^{(3)}$ leads to an appropriate increasing a total quantity of known Yin-Yang-matrices and their Hadamard genomatrices. We receive a further increase in this total quantity else, if we consider other initial groups of such genomatrices, which are transformed cyclically each into another at the same cyclic permutation of genetic elements C→A→G→U→C. Examples of such groups are $[C\ \ G;\ U\ \ A]^{(3)} \rightarrow [U\ \ C;\ A\ \ G]^{(3)} \rightarrow [A\ \ U;\ G\ \ C]^{(3)} \rightarrow [G\ \ A;\ C\ \ U]^{(3)} \rightarrow [C\ \ G;\ U\ \ A]^{(3)}$, or $[C\ \ A;\ G\ \ U]^{(3)} \rightarrow [G\ \ C;\ U\ \ A]^{(3)} \rightarrow [U\ \ G;\ A\ \ C]^{(3)} \rightarrow [A\ \ U;\ C\ \ G]^{(3)} \rightarrow [C\ \ A;\ G\ \ U]^{(3)}$, or $[G\ \ A;\ U\ \ C]^{(3)} \rightarrow [U\ \ G;\ C\ \ A]^{(3)} \rightarrow [C\ \ U;\ A\ \ G]^{(3)} \rightarrow [A\ \ C;\ G\ \ U]^{(3)} \rightarrow [G\ \ A;\ U\ \ C]^{(3)}$, or $[G\ \ C;\ A\ \ U]^{(3)} \rightarrow [A\ \ G;\ U\ \ C]^{(3)} \rightarrow [U\ \ A;\ C\ \ G]^{(3)} \rightarrow [C\ \ U;\ G\ \ A]^{(3)} \rightarrow [G\ \ C;\ A\ \ U]^{(3)}$. Such groups of matrices can be transformed each into other by means of permutations of separate genetic elements like C→G→C or A→G→A, etc.

We have no possibility to demonstrate in one article all hierarchy of cyclic sequences of genetic Yin-Yang-algebras and conjunct Hadamard matrices, which arise in the result of all possible kinds of cyclic permutations of genetic elements. Additional data about this hierarchy of "round dances" of Yin-Yang-algebras should be published later. An important general result is that a considerable quantity of genetic Yin-Yang-matrices and of their Hadamard matrices exists and that these matrices are connected with different kinds of cyclic permutations of genetic elements. Each of these Yin-Yang-matrices can be transformed into the initial matrix $YY_{+,123}^{[CAUG]}$ by means of a relevant cyclic permutation of genetic elements.

In addition, a set of such hierarchies of cyclic metamorphoses of the genetic Yin-Yang-algebras allows modeling of phenomena of metamorphoses of animals. For example, butterflies and moths have four stages of cyclic metamorphoses in their life: egg, larva (the caterpillar stage), pupa (the chrysalis phase), and adult. One should note that in the chrysalis phase the biological organism does not eat at all; consequently atomic contents of the organism do not change practically, but its set of molecular compositions is reformed at this stage cardinally by means of complex permutations of groups of chemical elements. In the proposed modeling approach, each of named stages of metamorphosis is connected with forming its own kind of hierarchy of the genetic Yin-Yan-algebras. Correspondingly a transition from one stage of biological metamorphosis to another stage is interpreted as a transition from one kind of hierarchy of the genetic Yin-Yang-algebras to another kind of their hierarchy. Some results of such modeling will be published separately.

### 3. The oppositional category of genetic Yin-Yang-matrices

Let us return to the genetic matrix algebra $YY_+^{[CAUG]}$ (Figures 1 and 2). If all its Yang-coordinates are equal to zero ($x_1 = x_3 = x_5 = x_7 = 0$), the $YY_+^{[CAUG]}$-matrix becomes the matrix of the genoquaternion of the Yin-type (Figure 14, on the left side). We mark this "female" genoquaternion by the symbol $G_f$. If all Yin-coordinates are equal to zero ($x_0 = x_2 = x_4 = x_6 = 0$), the $YY_+^{[CAUG]}$-matrix becomes the matrix of the genoquaternion of the Yang-type. We mark this "male" genoquaternion by the symbol $G_m$.

| | | | | | | | |
|---|---|---|---|---|---|---|---|
| $x_0$ | 0 | $-x_2$ | 0 | $x_4$ | 0 | $-x_6$ | 0 |
| $x_0$ | 0 | $-x_2$ | 0 | $x_4$ | 0 | $-x_6$ | 0 |
| $x_2$ | 0 | $x_0$ | 0 | $-x_6$ | 0 | $-x_4$ | 0 |
| $x_2$ | 0 | $x_0$ | 0 | $-x_6$ | 0 | $-x_4$ | 0 |
| $x_4$ | 0 | $-x_6$ | 0 | $x_0$ | 0 | $-x_2$ | 0 |
| $x_4$ | 0 | $-x_6$ | 0 | $x_0$ | 0 | $-x_2$ | 0 |
| $-x_6$ | 0 | $-x_4$ | 0 | $x_2$ | 0 | $x_0$ | 0 |
| $-x_6$ | 0 | $-x_4$ | 0 | $x_2$ | 0 | $x_0$ | 0 |
| 0 | $x_1$ | 0 | $-x_3$ | 0 | $x_5$ | 0 | $-x_7$ |
| 0 | $x_1$ | 0 | $-x_3$ | 0 | $x_5$ | 0 | $-x_7$ |
| 0 | $x_3$ | 0 | $x_1$ | 0 | $-x_7$ | 0 | $-x_5$ |
| 0 | $x_3$ | 0 | $x_1$ | 0 | $-x_7$ | 0 | $-x_5$ |
| 0 | $x_5$ | 0 | $-x_7$ | 0 | $x_1$ | 0 | $-x_3$ |
| 0 | $x_5$ | 0 | $-x_7$ | 0 | $x_1$ | 0 | $-x_3$ |
| 0 | $-x_7$ | 0 | $-x_5$ | 0 | $x_3$ | 0 | $x_1$ |
| 0 | $-x_7$ | 0 | $-x_5$ | 0 | $x_3$ | 0 | $x_1$ |

*Figure 14. On the left side: the matrix $G_f$ of Yin-genoquaternion. On the right side: the matrix $G_m$ of Yang-genoquaternion. Columns with null components are marked by green color. Matrix cells with positive (negative) components are marked by black (white) colors as in all previous cases.*

Multiplication tables of the algebras of these genoquaternions were shown in [Petoukhov, arXiv:0803.3330, Figures 13 and 14]. The sum of these two matrices $G_f$ and $G_m$ gives the $YY_+^{[CAUG]}$-matrix (Figures 1 and 2):

$$G_f + G_m = YY_+^{[CAUG]} \qquad (3)$$

The typical feature of this $YY_+^{[CAUG]}$-matrix and of all other Yin-Yang-matrices, which have been considered in the previous paragraph and which possesses the lower index "+", is that two halves of each of these matrices are mirror-antisymmetric in their black-and-white mosaics. For example, in the case of $YY_+^{[CAUG]}$-matrix (Figures 1 and 2) its left half and its right half are mirror-antisymmetric: each pair of cells, which are disposed mirror-symmetrically in these halves, have opposite colors. All these Yin-Yang-matrices can be produced by means of summation of relevant matrices of a Yin-qenoquaternion and of a Yang-qenoquaternion.

But what kind of a matrix arises in the case of subtraction of one genoquaternion from another genoquaternion? We mark the arisen matrix by a symbol $YY_-^{[CAUG]}$ with the lower index “-“:

$$G_f - G_m = YY_-^{[CAUG]} \quad (4)$$

| | | | | | | | |
|---|---|---|---|---|---|---|---|
| $x_0$ | $-x_1$ | $-x_2$ | $x_3$ | $x_4$ | $-x_5$ | $-x_6$ | $x_7$ |
| $x_0$ | $-x_1$ | $-x_2$ | $x_3$ | $x_4$ | $-x_5$ | $-x_6$ | $x_7$ |
| $x_2$ | $-x_3$ | $x_0$ | $-x_1$ | $-x_6$ | $x_7$ | $-x_4$ | $x_5$ |
| $x_2$ | $-x_3$ | $x_0$ | $-x_1$ | $-x_6$ | $x_7$ | $-x_4$ | $x_5$ |
| $x_4$ | $-x_5$ | $-x_6$ | $x_7$ | $x_0$ | $-x_1$ | $-x_2$ | $x_3$ |
| $x_4$ | $-x_5$ | $-x_6$ | $x_7$ | $x_0$ | $-x_1$ | $-x_2$ | $x_3$ |
| $-x_6$ | $x_7$ | $-x_4$ | $x_5$ | $x_2$ | $-x_3$ | $x_0$ | $-x_1$ |
| $-x_6$ | $x_7$ | $-x_4$ | $x_5$ | $x_2$ | $-x_3$ | $x_0$ | $-x_1$ |

*Figure 15. The matrix $YY_-^{[CAUG}$.*

Figure 15 shows this new matrix $YY_-^{[CAUG]}$. One can see, that the black-and-white mosaic of this matrix possesses unexpectedly a relation of mirror-symmetry between the left half and the right half in contrast to the case of $YY_+^{[CAUG]}$-matrix. Figure 16 shows tessellations of a plane by these mosaics for an additional comparison.

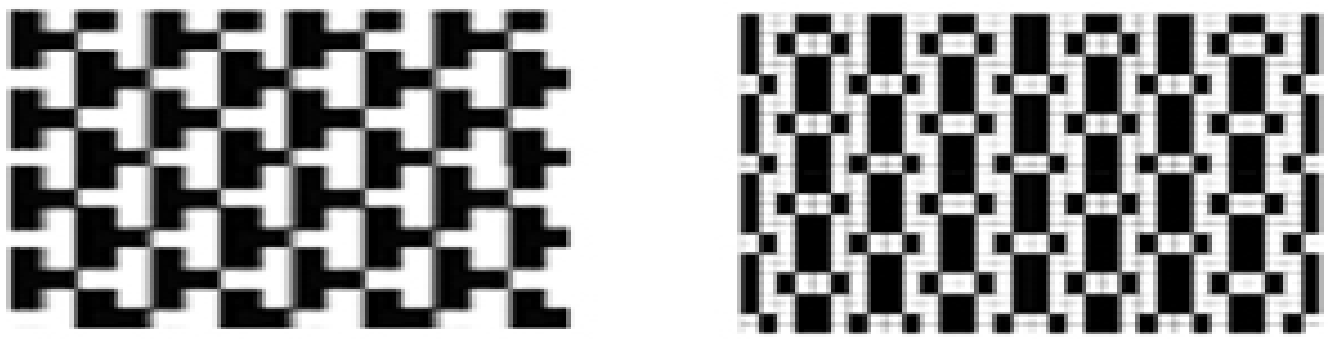

*Figure 16. The tessellations of a plane by the mosaic of the $YY_+^{[CAUG]}$-matrix (on the left side) and by the mosaic of the $YY_-^{[CAUG]}$-matrix (on the right side)*

The second unexpected fact is that this $YY_-^{[CAUG]}$-matrix defines its own Yin-Yang-algebra as well. The multiplication table of this $YY_-^{[CAUG]}$-algebra is shown on Figure 17.

| | $f_0$ | $m_1$ | $f_2$ | $m_3$ | $f_4$ | $m_5$ | $f_6$ | $m_7$ |
|---|---|---|---|---|---|---|---|---|
| $f_0$ | $f_0$ | $m_1$ | $f_2$ | $m_3$ | $f_4$ | $m_5$ | $f_6$ | $m_7$ |
| $m_1$ | $-f_0$ | $-m_1$ | $-f_2$ | $-m_3$ | $-f_4$ | $-m_5$ | $-f_6$ | $-m_7$ |
| $f_2$ | $f_2$ | $m_3$ | $-f_0$ | $-m_1$ | $-f_6$ | $-m_7$ | $f_4$ | $m_5$ |
| $m_3$ | $-f_2$ | $-m_3$ | $f_0$ | $m_1$ | $f_6$ | $m_7$ | $-f_4$ | $-m_5$ |
| $f_4$ | $f_4$ | $m_5$ | $f_6$ | $m_7$ | $f_0$ | $m_1$ | $f_2$ | $m_3$ |
| $m_5$ | $-f_4$ | $-m_5$ | $-f_6$ | $-m_7$ | $-f_0$ | $-m_1$ | $-f_2$ | $-m_3$ |
| $f_6$ | $f_6$ | $m_7$ | $-f_4$ | $-m_5$ | $-f_2$ | $-m_3$ | $f_0$ | $m_1$ |
| $m_7$ | $-f_6$ | $-m_7$ | $f_4$ | $m_5$ | $f_2$ | $m_3$ | $-f_0$ | $-m_1$ |

*Figure 17. The multiplication table of the genetic $YY_-^{[CAUG]}$-algebra.*

By analogy with the previous paragraph, one can analyze transformations of the genetic $YY_-^{[CAUG]}$-matrix, which are produced in the result of the same cyclic permutation of its genetic elements C→A→G→U→C. The same series of these cyclic permutations leads to arising a cyclic sequence of the following genetic matrices, which are shown on Figure 18: $YY_-^{[CAUG]}$ →$YY_-^{[UCGA]}$ →$YY_-^{[GUAC]}$ →$YY_-^{[AGCU]}$ →$YY_-^{[CAUG]}$. This Figure 18 is the analogue of the Figure 12 for the genetic Yin-Yang-matrices, which were described in the previous paragraph.

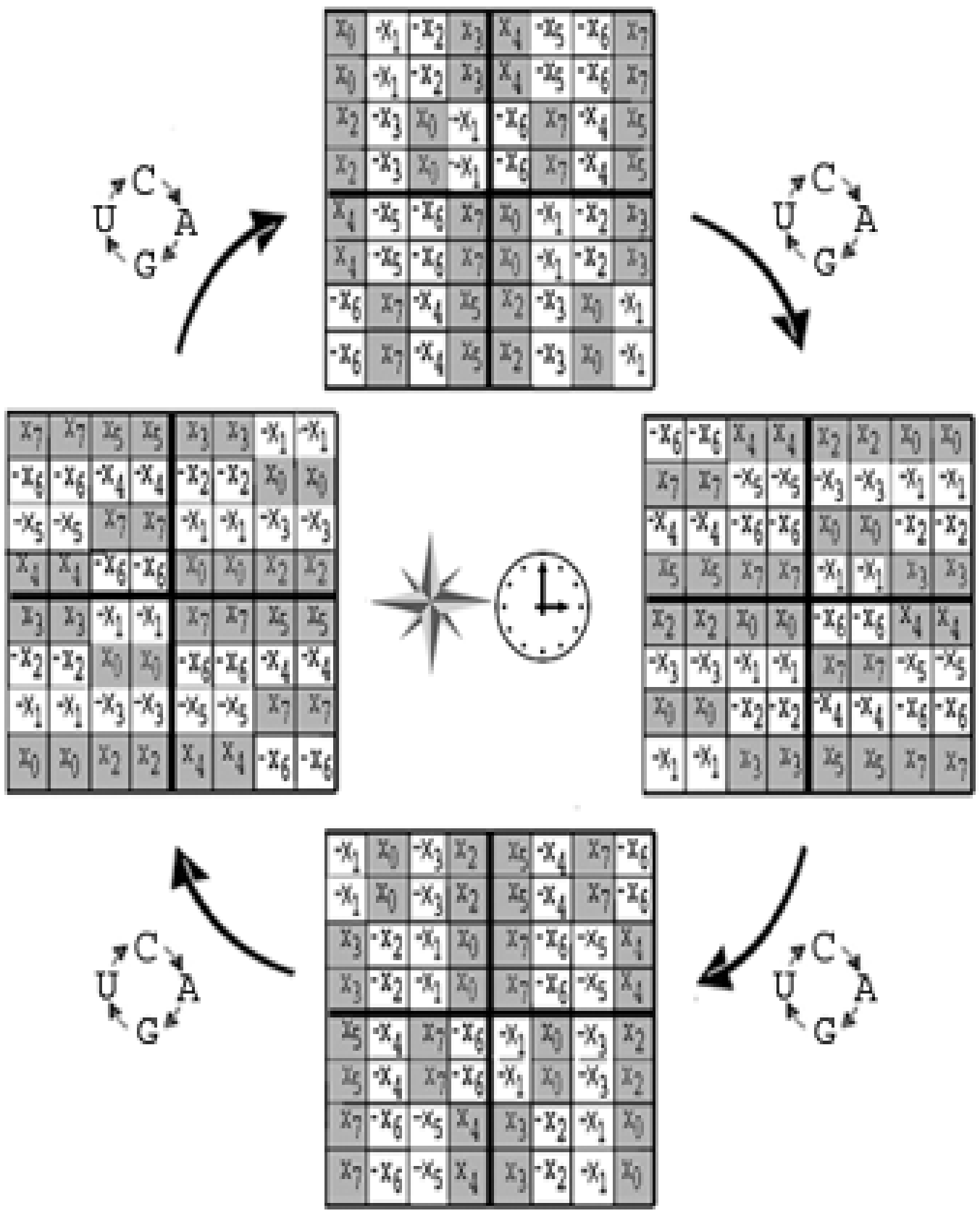

*Figure 18. The cyclic sequence of the Yin-Yang-matrices* $YY_-^{[CAUG]} \rightarrow YY_-^{[UCGA]} \rightarrow YY_-^{[GUAC]} \rightarrow YY_-^{[AGCU]} \rightarrow YY_-^{[CAUG]}$*, which arises on the bases of the cyclic permutation of the genetic elements C→A→G→U→C in the matrices.*

Each of these new three matrices $YY_-^{[UCGA]}$, $YY_-^{[GUAC]}$ and $YY_-^{[AGCU]}$ is the Yin-Yang-matrix, which defines its relevant Yin-Yang-algebra. The multiplication tables of basic elements of these new Yin-Yang-matrices are shown on Figure 19-21.

| | $f_0$ | $m_1$ | $f_2$ | $m_3$ | $f_4$ | $m_5$ | $f_6$ | $m_7$ |
|---|---|---|---|---|---|---|---|---|
| $f_0$ | $-f_6$ | $f_6$ | $-f_4$ | $f_4$ | $-f_2$ | $f_2$ | $-f_0$ | $f_0$ |
| $m_1$ | $-m_7$ | $m_7$ | $-m_5$ | $m_5$ | $-m_3$ | $m_3$ | $-m_1$ | $m_1$ |
| $f_2$ | $f_4$ | $-f_4$ | $-f_6$ | $f_6$ | $f_0$ | $-f_0$ | $-f_2$ | $f_2$ |
| $m_3$ | $m_5$ | -m5 | $-m_7$ | $m_7$ | $m_1$ | $-m_1$ | $-m_3$ | $m_3$ |
| $f_2$ | f2 | $-f_2$ | $-f_0$ | $f_0$ | $f_6$ | $-f_6$ | $-f_4$ | $f_4$ |
| m5 | $m_3$ | $-m_3$ | $-m_1$ | $m_1$ | $m_7$ | $-m_7$ | $-m_5$ | $m_5$ |
| $f_6$ | $-f_0$ | $f_0$ | $-f_2$ | $f_2$ | $-f_4$ | $f_4$ | $-f_6$ | $f_6$ |
| $m_7$ | $-m_1$ | $m_1$ | $-m_3$ | $m_3$ | $-m_5$ | $m_5$ | $-m_7$ | $m_7$ |

*Figure 19. The multiplication tables of the* $YY_-^{[UCGA]}$*-algebra*

| | $f_0$ | $m_1$ | $f_2$ | $m_3$ | $f_4$ | $m_5$ | $f_6$ | $m_7$ |
|---|---|---|---|---|---|---|---|---|
| $f_0$ | $f_0$ | $m_1$ | $f_2$ | $m_3$ | $f_4$ | $m_5$ | $f_6$ | $m_7$ |
| $m_1$ | $-f_0$ | $-m_1$ | $-f_2$ | $-m_3$ | $-f_4$ | $-m_5$ | $-f_6$ | $-m_7$ |
| $f_2$ | $f_2$ | $m_3$ | $-f_0$ | $-m_1$ | $-f_6$ | $-m_7$ | $f_4$ | $m_5$ |
| $m_3$ | $-f_2$ | $-m_3$ | $f_0$ | $m_1$ | $f_6$ | $m_7$ | $-f_4$ | $-m_5$ |
| $f_4$ | $f_4$ | $m_5$ | $f_6$ | $m_7$ | $f_0$ | $m_1$ | $f_2$ | $m_3$ |
| $m_5$ | $-f_4$ | $-m_5$ | $-f_6$ | $-m_7$ | $-f_0$ | $-m_1$ | $-f_2$ | $-m_3$ |
| $f_6$ | $f_6$ | $m_7$ | $-f_4$ | $-m_5$ | $-f_2$ | $-m_3$ | $f_0$ | $m_1$ |
| $m_7$ | $-f_6$ | $-m_7$ | $f_4$ | $m_5$ | $f_2$ | $m_3$ | $-f_0$ | $-m_1$ |

*Figure 20. The multiplication tables of the the $YY_{-}^{[GUAC]}$-algebra.*

| | $f_0$ | $m_1$ | $f_2$ | $m_3$ | $f_4$ | $m_5$ | $f_6$ | $m_7$ |
|---|---|---|---|---|---|---|---|---|
| $f_0$ | $-f_6$ | $f_6$ | $-f_4$ | $f_4$ | $-f_2$ | $f_2$ | $- f_0$ | $f_0$ |
| $m_1$ | $-m_7$ | $m_7$ | $-m_5$ | $m_5$ | $-m_3$ | $m_3$ | $-m_1$ | $m_1$ |
| $f_2$ | $f_4$ | $-f_4$ | $-f_6$ | $f_6$ | $f_0$ | $- f_0$ | $-f_2$ | $f_2$ |
| $m_3$ | $m_5$ | -m5 | $-m_7$ | $m_7$ | $m_1$ | $-m_1$ | $-m_3$ | $m_3$ |
| $f_2$ | f2 | $-f_2$ | $- f_0$ | $f_0$ | $f_6$ | $-f_6$ | $-f_4$ | $f_4$ |
| m5 | $m_3$ | $-m_3$ | $-m_1$ | $m_1$ | $m_7$ | $-m_7$ | $-m_5$ | $m_5$ |
| $f_6$ | $- f_0$ | $f_0$ | $-f_2$ | $f_2$ | $-f_4$ | $f_4$ | $-f_6$ | $f_6$ |
| $m_7$ | $-m_1$ | $m_1$ | $-m_3$ | $m_3$ | $-m_5$ | $m_5$ | $-m_7$ | $m_7$ |

*Figure 21. The multiplication tables of the $YY_{-}^{[AGCU]}$-algebra.*

All four genetic matrices $YY_{-}^{[CAUG]}$, $YY_{-}^{[UCGA]}$, $YY_{-}^{[GUAC]}$ and $YY_{-}^{[AGCU]}$ are connected with their own Hadamard matrices by means of the same algorithmic way (Figure 22).

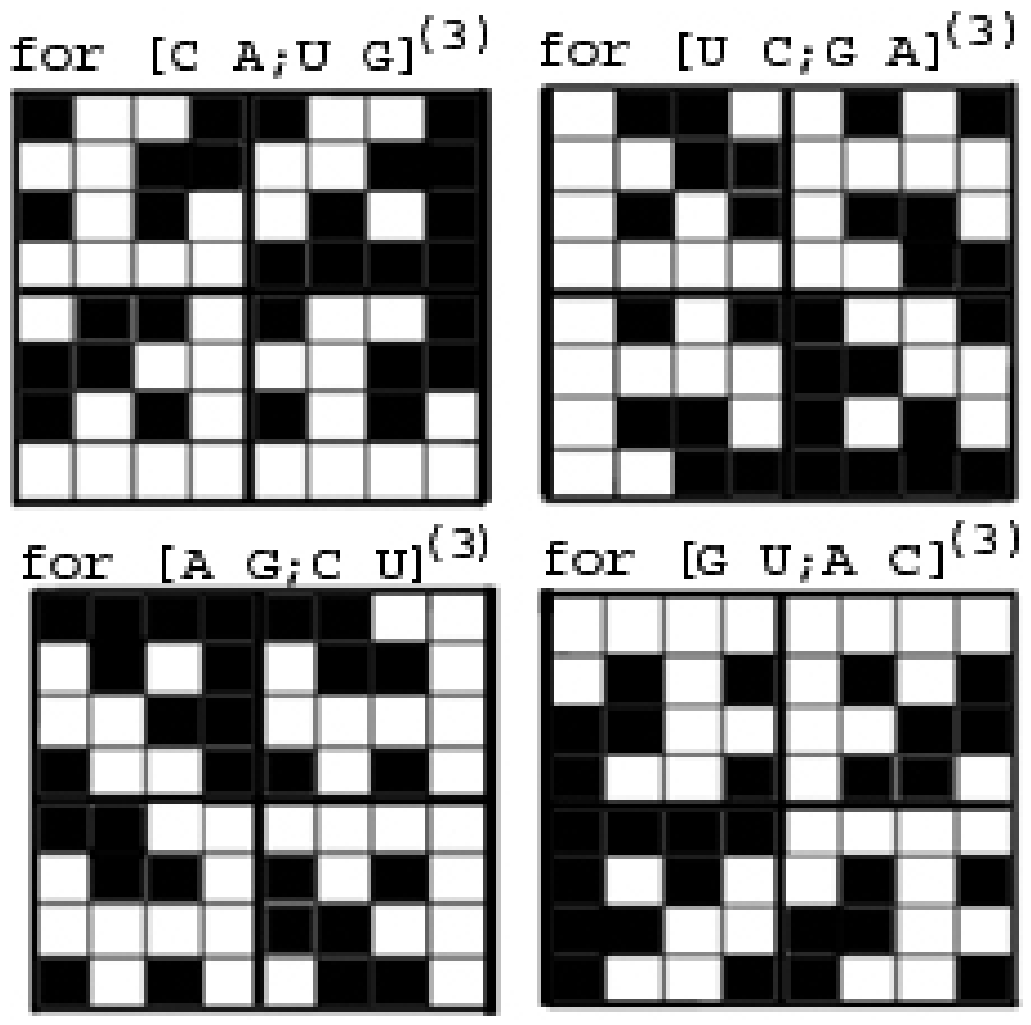

*Figure 22. The Hadamard matrices, which are connected algorithmically with the Yin-Yang-matrices $YY_{-}^{[CAUG]}$, $YY_{-}^{[UCGA]}$, $YY_{-}^{[GUAC]}$ and $YY_{-}^{[AGCU]}$. Each black (white) cell contains the element "+1" ("-1").*

By analogy with the previous paragraph, a total quantity of genetic Yin-Yang-matrices of this oppositional category (and a quantity of their relevant Hadamard matrices) increases significantly if one takes into a consideration the same described permutations of the genetic elements including the permutations of positions in triplets, etc. The author estimates preliminarily that a total quantity of Yin-Yang-matrices of the both categories exceeds 1000 matrices considerably. But the question about a precise quantity of such Yin-Yang-matrices, which form a hierarchy of genetic Yin-Yang-algebras, is one of new open questions of the matrix genetics.

## 4 Generative and suppressive properties of Yin-Yang-matrices of the two categories

The connection of described Yin-Yang-matrices with their relevant Hadamard matrices seems to be interesting and perspective for study. In this paragraph we will consider a case when magnitudes of all coordinates of the Yin-genoquaternion $G_f$ and of the Yang-quaternion $G_m$ (Figure 14) are equal to 1 ($x_0 = x_1 = x_2 = x_3 = x_4 = x_5 = x_6 = x_7 = 1$). We will name conditionally such genoquaternions and their mentioned combinations ($G_f + G_m$) and ($G_f - G_m$) as "elementary": the elementary Yin-genoquaternion, the elementary $YY_+^{[CAUG]}$-matrix, etc. This case is interesting specially because all components of Yin-Yang-matrices of the both oppositional categories ($YY_+^{[CAUG]}$, $YY_-^{[CAUG]}$, etc.) are equal in this case to "+1" or "-1" like in Hadamard matrices. Simple changes of the signs "+' or "-" of some components of such elementary Yin-Yang-matrices in accordance with the U-algorithm are enough to transform these elementary Yin-Yang-matrices into relevant Hadamard matrices.

The elementary genetic Yin-Yang-matrices of the two oppositional categories possess some beautiful properties relative to multiplication. If any elementary Yin-Yang-matrix ($G_f + G_m$) is raised into the second power, the result is a tetra-reproduction of this Yin-Yang-matrix. For example, let us consider such exponentiation of each of the six Yin-Yang-matrices on Figure 13:

$$(YY_{+,123}^{[CAUG]})^2 = 4* YY_{+,123}^{[CAUG]}; \quad (YY_{+,231}^{[CAUG]})^2 = 4* YY_{+,231}^{[CAUG]};$$
$$(YY_{+,312}^{[CAUG]})^2 = 4* YY_{+,312}^{[CAUG]}; \quad (YY_{+,321}^{[CAUG]})^2 = 4* YY_{+,321}^{[CAUG]};$$
$$(YY_{+,213}^{[CAUG]})^2 = 4* YY_{+,213}^{[CAUG]}; \quad (YY_{+,132}^{[CAUG]})^2 = 4* YY_{+,132}^{[CAUG]} \qquad (5)$$

This property can be illustrated graphically as arising four identical matrices instead of one initial matrix. It generates some associations with the tetra-reproduction of gametal cells in a process of meiosis. In view of this, we name elementary Yin-Yang-matrices ($G_f + G_m$) as generative Yin-Yang-matrices or start-matrices. By the way, if the elementary Yin-genoquaternion $G_f$ or the elementary Yang-genoquaternion $G_m$ is raised into the second power, the result is a double-reproduction of this elementary genoquaternion: $G_f^2 = 2* G_f$ and $G_m^2 = 2* G_m$. It reminds of a double-reproduction (a dichotomy) of somatic cells in a process of mitosis.

On the contrary, elementary Yin-Yang-matrices ($G_f - G_m$) possess a suppressive property: if any elementary Yin-Yang-matrix ($G_f - G_m$) is raised into the second power, the result is the null matrix. For comparison with the expression (5), one can show the following results:

$$(YY_{-,123}^{[CAUG]})^2 = 0; \quad (YY_{-,231}^{[CAUG]})^2 = 0;$$
$$(YY_{-,312}^{[CAUG]})^2 = 0; \quad (YY_{-,321}^{[CAUG]})^2 = 0;$$
$$(YY_{-,213}^{[CAUG]})^2 = 0; \quad (YY_{-,132}^{[CAUG]})^2 = 0 \qquad (6)$$

In view of this, we name elementary Yin-Yang-matrices ($G_f - G_m$) as suppressive Yin-Yang-matrices or stop-matrices (or apoptosis-matrices). A generative Yin-Yang-matrix can be transformed into a suppressive Yin-Yang-matrix and vice versa by means of the simple inversion of the signs "+" and "-" in Yang-components (or in Yin-components) of this Yin-Yang-matrix. Such transformations can be defined as functions of time by means of a definition of Yin-Yang-coordinates as relevant functions of time.

Many interesting relations exist among generative and suppressive Yin-Yang-matrices and genoquaternions $G_f$ and $G_m$. For example, product of a generative Yin-Yang-matrix with a suppressive Yin-Yang-matrix is equal to the tetra-reproduction of the suppressive Yin-Yang-matrix: $(YY_{+,123}^{[CAUG]}) * (YY_{-,123}^{[CAUG]}) = 4*(YY_{-,123}^{[CAUG]})$, etc. On the contrary, product of a suppressive matrix with a Yin-Yang-matrix ($G_f + G_m$) is equal to the null matrix. A set of such relations and properties gives new possibilities to create mathematical models of self-developing biological systems.

Whether molecular-genetic facts exist about an important role of cyclic or circular principles in genetic systems? Yes, interesting phenomenological results of studying some circular principles of organizations of molecular-genetic systems are presented, for example, in articles [Arques, Michel, 1996, 1997; Frey, Michel, 2003, 2006; Stambuk, 1999]. One can hope that our algebraic-genetic results and these published phenomenological results will supplement each other for deeper understanding the genetic systems.

Our results about hierarchy of cyclic sequences of cyclic changes of genetic Yin-Yang-algebras testify into favor of the following: a set of cyclic transformations and cyclic structures in living matter is connected with these cycles of algebraic changes or can be modeled by means of such changes of the genetic algebras. The nature has created the genetic code in such manner that a wide set of cyclic permutations of molecular-genetic elements in genetic matrices (or in these matrix forms of presentation of the genetic code) leads to cyclic changes of their genetic Yin-Yang-algebras. In author's opinion, a variety of species in living matter is connected in significant extent with the variety of types of genetic Yin-Yang-algebras and with the variety of cyclic permutations of the genetic elements. In particular, wide researches should be done about a natural division of all set of genes and proteins into special sub-sets by criteria of mutual cyclic transformations of members of each sub-set by means of cyclic permutations of their genetic elements (like the considered cyclic permutation C→A→G→U→C, which transforms any sequence of triplets into another sequence of triplets).

Whether algorithmic principles of organization of many cyclic movements (walking, run, breath, cardio cycles, etc.) of a separate individual can be presented as well in forms of mathematical models and algorithms, which are based on such changes of the Yin-Yang-algebras? In our opinion, it is possible in future. In such models, a transition of one kind of cyclic movements to another kind can be expressed as a transition from one cyclic sequence of changes of Yin-Yang-algebras to another cyclic sequence. One can think additionally about a possible relation of our algebraic-genetic approach with the famous conception by Eigen about hypercycles in biological organizations [Eigen, 1979].

## 5 Joining of the idea by Pythagoras and the idea of cyclic changes. About a notion of biological time.

Pythagoras has formulated the famous idea: "All things are numbers". Such known slogans of Pythagoreans as "numbers operate the world", "the world is number" reflect representations of Pythagoreans. For Pythagoreans the systems of numbers expressed "essence" of everything. In view of this idea, the natural phenomena should be explain by means of systems of numbers; the systems of numbers play a role of the beginning for uniting all things and for expressing the harmony of the nature [Kline,1980, p. 21, 24]. Many prominent scientists and thinkers were supporters of this viewpoint or of similar to it. Not without reason B. Russell (1945) noted that he did not know other person who would exert such influence on thinking of people as Pythagoras.

The history of science knows many thinkers who believed that all physics can be described on a language of some multi-dimensional numeric system or algebra. For example, W. Hamilton believed that all physics can be described on the language of his quaternions. This kind of thoughts belongs to a line of the Pythagorean idea. The data of matrix genetics about cyclic changes of the genetic Yin-Yang-algebras (or of the genetic Yin-Yang-numeric systems) give materials for another kind of thoughts or for a broadening the idea by Pythagoras. The speech is that new idea about organization of living matter arises. In accordance with this idea, organization of living matter is based not on a single algebra but on cyclic changes of many algebras of a certain set (a set of genetic Yin-Yang-algebras). It means that the idea by Pythagoras about numeric harmony of nature should be supplemented by another idea of cyclic changes of Yin-Yang-numeric systems. This new additional idea about cyclic changes of algebras in living matter reminds of the idea about cyclic changes from the Ancient Chinese "The Book of changes" ("I Ching") which was written about Yin-Yang-systems a few thousand years ago. But instead of the quite wide word "a cyclic change" we use the strict mathematical notion of "a change of one genetic Yin-Yan-algebra into another genetic Yin-Yang-algebra". Can this idea about cyclic changes of Yin-Yang-algebras be applied for inanimate matter in some extent as well? The future will show. The author does not know any other theory in the field of mathematical natural sciences which is based on cyclic changes of multi-dimensional algebras. It seems that the genetic code leads us to the new category of theories of mathematical natural sciences, which are based on a conception of cyclic changes inside a bunch of much multi-dimensional algebras depending on time and spatial features.

From the viewpoint of the proposed algebraic-cyclic conception about organization of living matter, the notion of "biological time" can be defined as a factor of a general coordinating (or of a general synchronization) of many cycles of changes of genetic Yin-Yang-algebras inside the hierarchy of these changes. If an organism is a hierarchy of cyclic changes of genetic Yin-Yang-algebras, such biological time is dispersed on all choruses of such cyclic processes of an organism. A dispersing of biological time along the whole organism reminds of a dispersing of feeling of music along the whole organism (our brain has not a special center of music, and music appeals to the whole organism [Weinberger, 2004]). The proposed algebraic-genetic approach to the problem of biological time reminds of the famous viewpoint [Whitrow, 1961] that biological time is internal time inside a spatial region of living matter (this spatial region is isolated in some extent from other regions of Universe).

**Acknowledgments**: Described researches were made by the author in the frame of a long-term cooperation between Russian and Hungarian Academies of Sciences and in the frame of programs of "International Society of Symmetry in Bioinformatics" (USA, http://polaris.nova.edu/MST/ISSB) and of "International Symmetry Association" (Hungary, http://symmetry.hu/). The author is grateful to Frolov K.V., Darvas G., Kappraff J., Ne'eman Y., He M., Kassandrov V.V., Smolianinov V.V., Vladimirov Y.S. for their support.